\documentclass[manuscript,screen]{acmart}

\AtBeginDocument{%
  \providecommand\BibTeX{{%
    \normalfont B\kern-0.5em{\scshape i\kern-0.25em b}\kern-0.8em\TeX}}}
    
\def\markup{0}
\if\markup1
\newcommand{\rv}[1]{{\leavevmode\color{blue}#1}}
\else
\newcommand{\rv}[1]{#1}
\fi

\usepackage{siunitx}
\usepackage{url}
\usepackage{soul}
\usepackage{graphicx}
\usepackage{textcomp}
\usepackage{float} 
\usepackage{subfigure}
\usepackage{color}
\usepackage{multirow}
\usepackage{svg}
\usepackage{makecell}

\begin{document}

\title[How Can Haptic Feedback Assist People with BLV]{How Can Haptic Feedback Assist People with Blind and Low Vision (BLV): A Systematic Literature Review}

\author{Chutian Jiang}
\authornote{Both authors contributed equally to this research.}
\affiliation{%
%\institution{Computational Media and Arts Thrust}
\institution{The Hong Kong University of Science and Technology (Guangzhou)}
\city{Guangzhou}
\country{China}
}
\email{cjiang893@connect.hkust-gz.edu.cn}

\author{Emily Kuang}
\authornotemark[1]
\affiliation{%
%\institution{Golisano College of Computing and Information Sciences}
\institution{Rochester Institute of Technology}
\city{Rochester}
\country{USA}
}
\email{ek8093@rit.edu}

% \author{Pan Hui}
% \authornote{Co-supervisor}
% \affiliation{%
% \institution{Computational Media and Arts Thrust}
%   \institution{The Hong Kong University of Science and Technology (Guangzhou)}
%   \city{Guangzhou}
%   \country{China}
% }
% \affiliation{%
%   \institution{The Hong Kong University of Science and Technology}
%   \city{Hong Kong SAR}
%   \country{China}
% }
% \email{panhui@ust.hk}

\author{Mingming Fan}
\authornote{Corresponding author}
\affiliation{%
%\institution{Computational Media and Arts Thrust}
  \institution{The Hong Kong University of Science and Technology (Guangzhou)}
  \city{Guangzhou}
  \country{China}
}
\affiliation{%
  \institution{The Hong Kong University of Science and Technology}
  \city{Hong Kong SAR}
  \country{China}
}
\email{mingmingfan@ust.hk}
\renewcommand{\shortauthors}{Jiang, Kuang, and Fan}

\begin{abstract}
People who are blind or have low vision (BLV) encounter numerous challenges in their daily lives and work. To support them, various haptic assistive tools have been developed. Despite these advancements, the effective utilization of these tools---including the optimal haptic feedback and on-body stimulation positions for different tasks along with their limitations---remains poorly understood. Recognizing these gaps, we conducted a systematic literature review spanning two decades (2004-2024) to evaluate the development of haptic assistive tools within the HCI community. Our findings reveal that these tools are primarily used for understanding graphical information, providing guidance and navigation, and facilitating education and training, among other life and work tasks. We identified three main limitations: hardware limitations, functionality limitations, and UX and evaluation methods limitations. Based on these insights, we discuss potential research avenues and offer suggestions for enhancing the effectiveness of future haptic assistive technologies.
\end{abstract}




\maketitle

\section{Introduction}
The World Health Organization reported that around 2.2 billion people worldwide experienced blindness or low vision (BLV) in 2024 \cite{world2024world}. 
These individuals face substantial barriers in various areas, including employment, education, navigation, and daily activities, which can exacerbate the information gap and lead to inequality in accessing information \cite{brooks2014relationship, Kim2021, abu2010multimodal}. 
To mitigate these challenges, various assistive tools, particularly those utilizing audio and haptic technologies, have been developed. 
Haptic assistive tools, for example, provide feedback through touch and motion stimulation at various body positions such as the fingers, hands, wrists, waist, ankles, arms, shoulders, and feet \cite{ujitoko2021survey, csapo2015survey}. 
Compared to audio feedback, haptic feedback offers a more intuitive means of understanding graphical information and is increasingly being employed to aid people with BLV in their daily lives and work \cite{jiang2023understanding}. 
This work aims to examine how different haptic assistive tools---along with the types of haptic feedback and on-body stimulation positions used---support the needs of users with BLV.

Researchers have developed a diverse range of haptic assistive tools to enhance the lives and work of people with BLV \cite{bhowmick2017insight}.
These tools include tactile graphics/maps \cite{fusco2015tactile,melfi2020understanding}, refreshable braille displays/pin arrays \cite{hu2022smart,holloway2022animations}, tablet/smartphone integrated haptic actuators \cite{yoo2022perception,chu2022comparative}, haptic mice \cite{brayda2015the,brayda2013predicting}, haptic gloves \cite{soviak2016tactile,quek2013enabling}, haptic sliders \cite{fan2022slide,gay2021f2t}, robot/drone haptic devices \cite{huppert2021guidecopter,rahman2023take}, haptic bands \cite{kayhan2022a,lee2023novel}, 3D models \cite{lee2023tacnote, sargsyan20233D}, white canes \cite{nasser2020thermalcane,tanabe2021identification}, robotic-arm haptic devices \cite{espinosa2021virtual,lieb2020haptic}, and other handheld haptic devices \cite{liu2021tactile,Morelli2010vi}. 
Prior literature reviews have categorized these innovations from different perspectives, such as the type of haptic assistive tools \cite{ozioko2022smart,bhatnager2023analysis}, the tasks addressed \cite{kreimeier2020two,masal2023development}, and specific user demographics \cite{ahmed2018assistive,fardan2023systematic}. 
However, comprehensive analyses that consider how different types of \textbf{assistive tools} (categorized by form factors, such as white canes and refreshable braille displays/pin arrays), \textbf{haptic feedback} (classified by mechanisms, such as thermal feedback induced by stimulating free nerve endings and Krause end bulbs), and \textbf{on-body stimulation positions} (classified by body locations, such as the wrist and fingers) are selected and utilized for various BLV-related tasks are scarce. 
We selected these three attributes because they are fundamental design components that substantially affect the effectiveness of addressing challenges in various tasks.
% Such perspectives are crucial as they highlight the intricate relationship between the tools' designs and their targeted tasks, revealing a strong interdependence. 

The device's form factor impacts both the range of tasks it can perform and its usability. For instance, a heavy or bulky device may be unusable for precision tasks, leading to task failure.
Haptic feedback and on-body stimulation positions impact how efficiently information is conveyed. For example, mapping distinct haptic patterns to different pieces of information can improve sensemaking for people with BLV. 
Additionally, the distribution of haptic receptors varies across the body, influencing how much information can be effectively conveyed at different body locations \cite{johansson1979tactile}.

Additionally, the limitations of existing haptic assistive tools, their user experience (UX), and the evaluation methods for studying these tools have not been thoroughly investigated. 
Without a deep understanding of these aspects, researchers risk overlooking gaps that could inform the development of future tools and improvements to their UX and evaluation method, potentially missing opportunities to make meaningful contributions to the BLV community. 

Therefore, in this paper, we investigate the following two research questions (RQs):

\textbf{RQ1: How can different haptic assistive tools, haptic feedback, and on-body stimulation positions be utilized to assist people with BLV in various tasks?}

\textbf{RQ2: What are the limitations of current haptic assistive tools in terms of UX and evaluation method?}

To answer these RQs, we conducted a systematic literature review covering twenty years (2004-2024) of developments in haptic assistive tools within the HCI community. 
Our review involved 132 papers from eleven key conferences and journals, including CHI, UIST, ASSETS, Ubicomp/IMWUT, TACCESS, TOCHI, IEEE Access, IEEE WHC, IEEE Haptics, IEEE ToH, and IEEE TVCG. 
From this survey, we found that haptic assistive tools, haptic feedback, and on-body stimulation positions predominantly support individuals with BLV in four main areas: graphical information understanding, guidance and navigation, education and training, and other life and work-related tasks.

When examining different haptic assistive tools, we identified 14 types of haptic assistive tools: tactile graphics/maps, refreshable braille displays/pin arrays, robotic-arm haptic devices, haptic bands, 3D models, white canes, other handheld haptic devices, haptic mice, robot/drone haptic devices, tablet/smartphone integrated haptic actuators, haptic gloves, haptic sliders, mid-air haptic devices, and electrotactile devices. 
Our review found that most studies focused on graphical information understanding (58 out of 132 papers), followed by education/training (28/132), other life and work tasks (26/132), and guidance/navigation (20/132).
% using tactile graphics/maps to assist in graphical information understanding (22 out of 132 papers). 
% For guidance and navigation tasks, haptic bands were most frequently used (10/132). 
% Education and training were predominantly supported by tactile graphics/maps (5/132) and robotic-arm haptic devices (5/132). 
% Additionally, refreshable braille displays/pin arrays were most commonly utilized to assist with various other life and work tasks (8/132). 

One mid-air haptic device was specifically used for driving, where it conveyed street intersection layouts by applying varying pressures to the palm to indicate the directions of different roads ahead of the vehicle \cite{fink2023autonomous}. 
While our review found that mid-air haptic devices are rarely employed as haptic assistive tools, they can be used in novel applications for public assistive tools, as they require no direct contact and allow for effortless remote interaction \cite{paneva2020haptiread}.
% Specifically, one mid-air haptic device was used for other life and work tasks, specifically for driving, where it presented the map of a street intersection by applying different pressures on the palm to create the direction cues of different roads in front of the vehicle \cite{fink2023autonomous}. 
% Our review indicated that mid-air haptic devices have seldom been utilized as haptic assistive tools. 
% This work highlighted a novel application area for these devices, particularly in public assistive tools, as they require no direct contact and facilitate easy remote interaction for users \cite{paneva2020haptiread}.

In our analysis of haptic feedback, we identified pressure, kinesthetic feedback, vibration, skin-stretch, and thermal feedback as the primary types. 
Furthermore, a single haptic assistive tool can incorporate multiple types of haptic feedback. 
Our review revealed preferences for specific feedback mechanisms to support different tasks. 
Pressure-based feedback emerged as the most frequently reported mechanism in aiding graphical information understanding (43/132). 
For guidance and navigation tasks, both kinesthetic feedback and vibration were equally utilized, each mentioned in 14 studies. 
Education and training tasks, along with other life and work-related tasks, were primarily supported by pressure feedback, mentioned in 14 and 17 studies, respectively. 
One study specifically employed thermal feedback to enhance guidance and navigation for individuals with BLV by varying the direction and temperature, illustrating an innovative application of thermal feedback in haptic devices.

For the on-body stimulation positions used by haptic devices, we observed that fingers (101/132) and hands (37/132) were the most commonly targeted areas. 
Graphical information understanding, education/training, and other life and work tasks were mostly supported by stimulating fingers, while hands were mostly used for guidance/navigation.
Compared to these two areas, wrists, waist, ankles, arms, shoulders, feet, and heads were utilized in fewer than five papers. 
These include the work of Kaul et al., who used haptic cues around the heads of individuals with BLV for guidance and navigation \cite{Kaul2021around}, and Xu et al., who applied haptic cues to the users with BLV' shoulders, wrists, and ankles for similar purposes \cite{xu2020virtual}. 

Additionally, by clustering the various limitations of haptic assistive tools thematically, we synthesized the key challenges into three main areas: hardware limitations (including environmental effects on sensors and actuators, conspicuous appearance and noise, and cumbersome and heavy devices), functionality limitations (including lack of information display control, lack of detailed information, and few presentation modalities), and UX and evaluation method limitations (including lack of user group diversity consideration, lack of customization, limitations of the user study, and haptic device distrust). 
Based on our findings, we identified potential research directions by pinpointing tasks where the application of haptic assistive tools, feedback types, and on-body stimulation positions has been limited. 
We also proposed targeted improvements to enhance the effectiveness of current haptic assistive technologies.

In sum, our literature review yields two key contributions: 

\begin{itemize}
\item We analyzed and discussed previous works to understand the applications of various haptic assistive tools, haptic feedback, and on-body stimulation positions in tasks performed by people with BLV.
\item We identified the limitations in the design and evaluation of current haptic assistive tools and highlighted future directions for improvement.
\end{itemize}

\section{Related Work} 
We first summarize people with BLV's accessibility gaps in daily lives and work. 
Then, we discuss haptic assistive tools to assist them in various tasks.

\subsection{People with BLV's Accessibility Gaps In Daily Lives and Work}
The lack of access to visual information is considered one of the biggest challenges constraining people with BLV's independence and productivity \cite{gorlewicz2018graphical}. 
Recent studies have examined daily accessibility gaps for people with BLV, identifying specific difficulties in graphical information understanding \cite{fan2023accessibility,sharif2021understanding,holloway2020non,marriott2021inclusive,kim2023exploring,morash2015guiding,jung2021communicating,kim2023explain,zhan2024virtuwander,huang2023understanding,rong2022it}, guidance/navigation \cite{lee2022apply,LOW2020137,avila2015drone,nicholson2009shoptalk,williams2014just,abdolrahmani2017embracing,jiang2023understanding}, education/training \cite{cattaneo2011blind,zebehazy2014quality,brule2016mapsense}, and other aspects of daily life and work\cite{lei2022ishake,huang2023understanding,Jung2021ThroughHand,rong2022it}. 
These challenges encompass both technical and attitudinal barriers.
For graphical information understanding, specific obstacles include low efficiency in comprehending overviews, difficulty in extracting detailed information, and high cognitive load on short-term memory\cite{sharif2018evographs,fan2023accessibility}. 
For guidance/navigation, challenges such as disorientation, difficulty avoiding obstacles and crowds, risks associated with crossing streets, and the induction of social stigma are prevalent \cite{riazi2016outdoor,dos2022aesthetics}.
For education/training, people with BLV face difficulty with following blackboard and slide presentations, communicating effectively with sighted classmates, efficiently learning graphical content, creating new content, and overcoming social isolation \cite{de_silva2023understanding,oliveira2012the,zhang2023understanding,fanshawe2023enablers}. 
% The other tasks' specific challenges were also reported in prior works \cite{lei2022ishake,huang2023understanding,Jung2021ThroughHand,rong2022it}. 
Challenges in other daily life and work tasks have also been documented \cite{lei2022ishake,huang2023understanding,Jung2021ThroughHand,rong2022it}.
For example, Fan et al. found that people with BLV expressed concerns regarding promptly accessing local COVID-19 data and charts. 
Bar charts, line charts, and maps were rated as mostly inaccessible, while bubble charts and pie charts were inaccessible \cite{fan2023accessibility}. 
Sharif found that screen reader users with BLV spent significantly more time and were less accurate in extracting information than sighted people when interacting with digital data visualization \cite{sharif2021understanding}.
% They found that all participants had experienced some outdoor difficulties, such as incorrectly installing tactile ground surface indicators, obstacles on sidewalks, disorientation, fear of falling, etc \cite{riazi2016outdoor}. 
Riazi et al.'s semi-structured interviews revealed that all participants encountered outdoor difficulties, including incorrectly installed tactile ground surface indicators, sidewalk obstacles, disorientation, and fear of falling \cite{riazi2016outdoor}.
Similarly, De Silva et al. identified key challenges in students with BLV's body movement education and proposed potential solutions \cite{de_silva2023understanding}. 
Rong et al.'s study on BLV live streaming highlighted algorithmic biases that suppressed content creators with BLV  \cite{rong2022it}.  
To address the challenges mentioned above in graphical information understanding, guidance/navigation, education/training, and other aspects of daily life and work, a range of haptic assistive tools have been developed. 
These solutions are discussed in the following section.

\subsection{Haptic Assistive Tools}
Based on existing categorizations in literature as well as our review of haptic tools, we derived 14 categories: 1) tactile graphics/maps \cite{baker2014tactile,fusco2015tactile,melfi2020understanding,taher2015exploring,goncu2010usability}, 2) refreshable Braille displays/pin arrays \cite{leithinger2015shape,hu2022smart,Graphical2016,holloway2022animations,ohshima2021development,kobayashi2018basic,wall2006tac}, 3) tablet/smartphone integrated haptic actuators \cite{palani2017principles,Toennies2011toward,Rantala2009methods,yoo2022perception,chu2022comparative}
, 4) haptic mice \cite{brayda2015the,kim2011handscope,Headley2011roughness,brayda2013predicting,Strachan2013vipong,levesque2012adaptive,pietrzak2009creating}, 5) haptic gloves \cite{soviak2016tactile,quek2013enabling,oliveira2012the}, 6) haptic sliders \cite{fan2022slide,Tanaka2016haptic,gay2021f2t}, 7) robot/drone haptic devices \cite{huppert2021guidecopter,guinness2019robo,rahman2023take,kayukawa2020guiding}, 8) haptic bands \cite{kayhan2022a,lee2023novel,Kaul2021around,erp2020tactile,hirano2019synchronized}, 9) 3D models \cite{lee2023tacnote, sargsyan20233D,Nagassa20233D,brule2021beyond}, 10) white canes \cite{siu2020virtual,nasser2020thermalcane,tanabe2021identification,zhao2018enabling}, 11) robotic-arm haptic devices \cite{abu2010multimodal,espinosa2021virtual,lieb2020haptic,zhang2017multimodal}, 12) other handheld haptic devices \cite{liu2021tactile,sanchez2010usability,amemiya2009haptic,Morelli2010vi}, 13) mid-air haptic devices \cite{fink2023autonomous}, and electrotactile devices \cite{jiang2024designing,yoo2022perception}.

Prior work has reviewed haptic assistive tools from various perspectives, such as types of haptic assistive tools \cite{o2015designing,wabinski2019automatic,Ducasse2018,gorlewicz2020design,yang2021survey,Emily2017review,chouvardas2005tactile,ozioko2022smart,ruxandra2020wearable,bhatnager2023analysis}, different tasks \cite{ahmed2018assistive,liu2006survey,kreimeier2020two,masal2023development,vidal2007graphical,real2019navigation}, and specific user groups \cite{ahmed2018assistive,kim2013elicitation,fardan2023systematic}. 
For example, Yang et al. reviewed papers on tactile displays and refreshable Braille displays, focusing on their mechanisms and technical issues and making design suggestions from their results \cite{yang2021survey}. 
Ahmed et al. reviewed papers for students' mathematics education and recommended developing novel assistive learning technologies based on haptic and audio feedback for all students with BLV in mathematics learning \cite{ahmed2018assistive}. 
However, prior works either focused on a specific type of haptic assistive tool, a specific task, or a specific user group, and few have considered a holistic overview of how various haptic assistive tools, haptic feedback, and on-body stimulation positions are utilized across multiple tasks for people with BLV.
Our focus on these perspectives is motivated by their direct relevance to BLV tasks and their interconnected nature.
In particular, various haptic assistive tools are designed to solve problems in different BLV tasks, such as white canes for guidance/navigation and tactile graphics/maps for understanding graphical information. 
Additionally, various types of haptic feedback are chosen based on their efficiency in conveying specific types of information.  
% For example, vibration and kinesthetic feedback can efficiently convey navigation and notification cues, while pressures can better convey graphical information. 
For example, vibration and kinesthetic feedback can efficiently convey navigation and notification cues, while pressure is proficient for conveying graphical information \cite{see2022touch}.
Various on-body stimulation positions have different haptic perception thresholds and thus can be used for multiple tasks. 
For example, fingers are more sensitive to haptic stimuli than hands or wrists, enabling them to perceive finer details \cite{koo2016two,mancini2014whole}. 
In addition, we considered other perspectives, such as form factor designs and qualitative study designs. 
Moreover, few have highlighted the limitations of various haptic assistive tools and their UX and evaluation methods. 
Lack of such understanding may induce an overlook of potential research gaps in future haptic assistive tools development and UX and evaluation method design. 
Therefore, our work investigates how different haptic assistive tools, haptic feedback, and on-body stimulation positions have been used to support people with BLV with common tasks, and the limitations of haptic assistive tools and UX and evaluation methods.
\section{Method}
\begin{figure} [tbh!]
    \centering
    \includegraphics[scale=0.45]{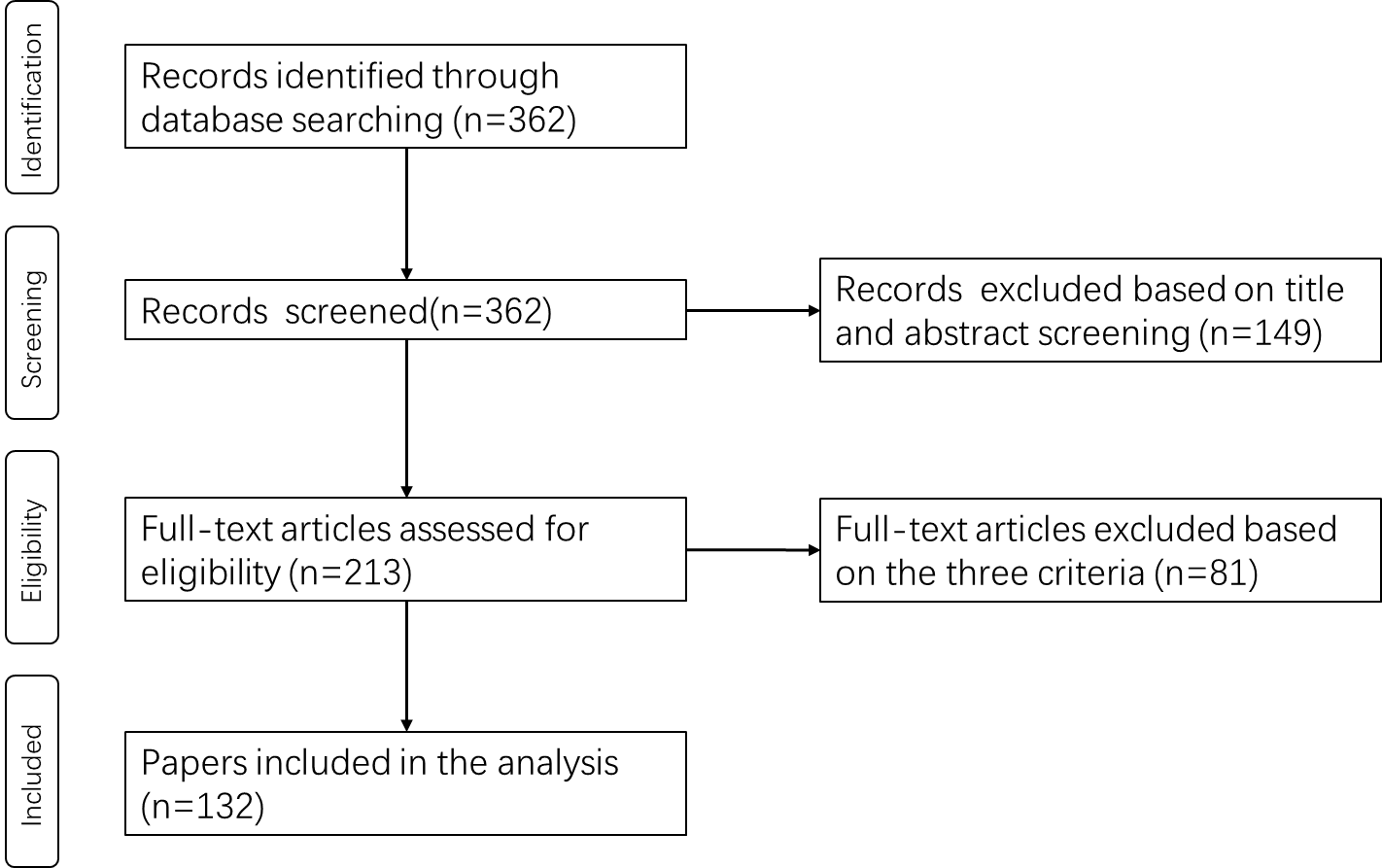}
     \caption{Four steps of literature review's PRISMA procedure: identification, screening, eligibility, and included.}
     \Description{This figure illustrates the PRISM procedure, which is a literature review method. The procedure includes four steps: identification, screening, eligibility, and included}
    \label{fig:PRISMA}
\end{figure}

We conducted a systematic literature review over the past twenty years to examine how haptic assistive tools assisted people with BLV. 
Following the PRISMA (Preferred Reporting Items for Systematic Reviews and Meta-Analyses) guidelines, which outline a four-phase procedure for systematic reviews, we selected papers for inclusion based on predefined criteria \cite{moher2009preferred}. 
These criteria were established to ensure the relevance and quality of the selected papers to our research question. 
The PRISMA procedure is illustrated in Figure \ref{fig:PRISMA}. 
In the following section, we provide a detailed overview of our search process.

To meet our research goal, we established three criteria to filter the literature:

\begin{enumerate}
\item \textbf{Involvement of individuals with BLV.}
Papers must include people with BLV, including those who are totally blind or have low vision.

\item \textbf{Focus on Assistive Technology.}
Selected papers must discuss assistive technology, defined as any tool, equipment, software, or system designed to enhance the functional abilities of people with disabilities \cite{wiat}. 

\item \textbf{Relevance to Haptic Devices.}
Papers should be directly related to haptic devices, which offer haptic experiences by applying forces, vibrations, or movements to the user's body.
\end{enumerate}

\subsection{Phase 1: Identifcation}
We aimed to identify high-impact papers on how haptic feedback is evaluated and published in venues on VR and HCI. Using Google Scholar Metrics, we identified eleven such venues: The ACM Conference on Human Factors in Computing Systems (CHI); The ACM Symposium on User Interface Software and Technology (UIST); The ACM SIGACCESS Conference on Computers and Accessibility (ASSETS); The ACM international joint conference on pervasive and ubiquitous computing (UbiComp) and The Proceedings of the ACM on Interactive, Mobile, Wearable and Ubiquitous Technologies (IMWUT); ACM Transactions on Accessible Computing (TACCESS); ACM Transactions on Computer-Human Interaction (TOCHI); IEEE Access; The World Haptics Conference (WHC); IEEE International Symposium on Haptic Interfaces for Virtual Environment and Teleoperator Systems (HAPTICS); IEEE Transactions on Haptics (ToH), IEEE Transactions on Visualization and Computer Graphics (TVCG).

To focus our work on haptic assistive tools for users with BLV, we used terms such as "haptic," "tactile," "blind," "low vision,""visually impaired," and "visual impairment." An example query of this on the ACM Digital Library is:\\

\textbf{Title: ((hapti* OR tactile) AND (blind OR low vision OR visually impaired OR visual impairment)) OR
Abstract: ((hapti* OR tactile) AND (blind OR low vision OR visually impaired OR visual impairment)) 
}\\

Where * denotes any number of unknown characters (wild cards). This way, we included words such as ‘haptic’ and ‘haptics’. The search approach differed across databases due to their unique search functionalities.
For IEEE venues, we listed the internal conference identifiers specific to Access, WHC, HAPTICS, ToH, and TVCG and added them to the query to include only those venues. 
And for CHI, UIST, Ubicomp/IMWUT, TACCESS, and TOCHI, we included the entire conference proceedings and excluded non-full papers (such as abstracts, posters, and other adjunct publications) in the latter phases.

Our search included full papers published in English over the past two decades (2004–2024). 
We focused on full papers because posters or adjunct publications often cannot provide the level of detail (due to limited paper length) about the experimental methodologies and results. 
We selected two decades to cover publications with most of the modern technologies for haptic assistive tools, yet resulted in an extensive set of papers to learn from. 
This resulted in 362 results: 153 from CHI, 12 from UIST, 51 from ASSETS, 11 from Ubicomp/IMWUT, 28 from TACCESS, 13 from TOCHI, 17 from Access, 21 from WHC, 16 from HAPTICS, 40 from ToH, and four from TVCG. We compiled the titles and abstracts of these 362 publications for screening in Phase 2.

\subsection{Phase 2: Screening}
We screened the titles and abstracts of the 362 papers collected in Phase 1 using the abovementioned inclusion criteria. 
For example, papers that did not mention the development of any haptic assistive tools for people with BLV were excluded. 
Two authors individually rated the same set of 20 randomly chosen papers for inclusion, resulting in Cohen’s Kappa of 0.84. 
Then, they rated the rest of the papers for inclusion. 
Out of the 362, we excluded 149 papers, resulting in 213 papers for phase 3.

\subsection{Phase 3: Eligibility}
We screened the full-text articles for eligibility with the three criteria. 
The reasons for exclusion in this phase were either (a) that a paper did not meet one or more of the three inclusion criteria despite the abstract screening, or (b) that a paper was not a full paper. 
In this phase, we excluded a further 81 publications. Thus, 132 papers were analyzed for the last phase. 

\subsection{Phase 4: Data Set and Coding Process}
The remaining 132 publications were included in the review. 
We coded each paper based on the haptic assistive tools, haptic feedback, and on-body stimulation positions, and tallied the frequencies of each code to determine the distribution of haptic assistive tools, haptic feedback, and on-body stimulation positions that assist in different tasks.

\subsection{Data Analysis}
In addition to categorizing the papers by task type, haptic feedback type, and stimulation locations, we conducted a comparative analysis of the different haptic tools, feedback methods, and stimulation positions to identify underexplored areas and potential opportunities.
Our comparative analysis involved contrasting the design, functionality, and application contexts of each haptic tool. 
We evaluated how different feedback methods---such as vibration, kinesthetic, and thermal feedback---were implemented across various systems. 
We categorized the types of haptic feedback based on the induction mechanism and the hardware described in each paper (e.g., kinesthetic feedback involves external forces like dragging, while vibration is generated by vibrators). 
We then divided the types of feedback by task to identify which were more or less commonly used for each task.
Additionally, we analyzed the stimulation positions (e.g., hand, arm, or full-body feedback) to assess their effectiveness in enhancing user interaction and task performance.
We also synthesized the overall limitations of these tools (Section \ref{Limitations of Haptic Assistive Tools and User Studies}). 
For each paper, we reviewed the limitations section when available. 
We then applied thematic clustering \cite{charmaz_constructing_2006}, using inductive coding to generate a list of limitations, which we grouped into subcategories and broader categories. 
These subcategories and categories were further refined through discussions among the authors.
We discussed possible improvements to address each limitation, resulting in our list of recommended improvements (Section \ref{Areas of Improvement}).

\subsection{Overview of Findings}
\begin{figure} [tbh!]
    \centering
    \includegraphics[scale=0.45]{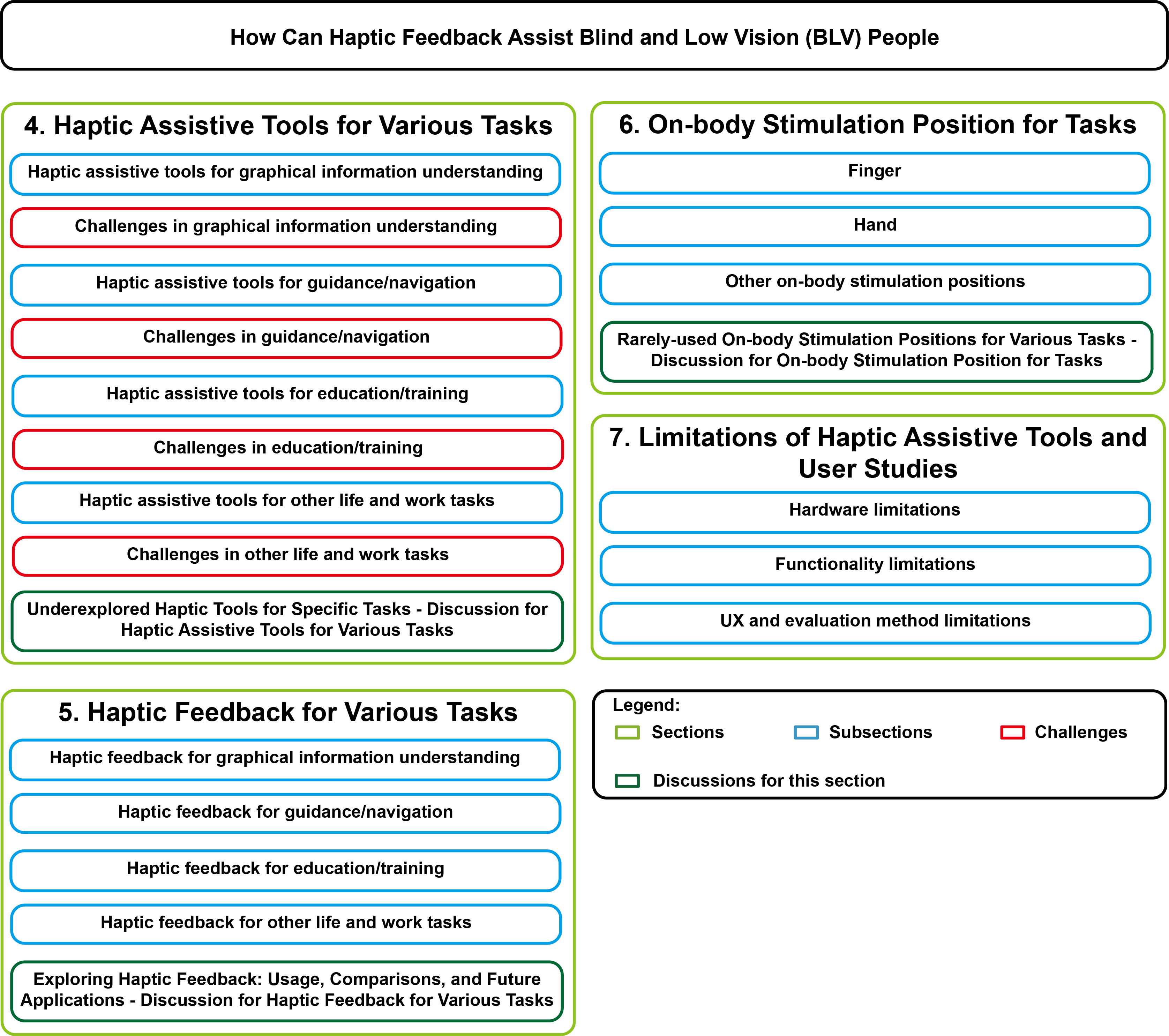}
     \caption{Findings reported from three main perspectives and limitations. Each section was further divided into subsections.}
     \Description{This figure illustrates the findings of three main perspectives and Limitations. Each section was further divided into subsections.}
    \label{fig:findings}
\end{figure}

In the following sections, we report how various \textbf{haptic assistive tools} (Section \ref{Haptic Assistive Tools for Various Tasks}), \textbf{haptic feedback} (Section \ref{Haptic Feedback for Various Tasks}), and \textbf{on-body stimulation positions} (Section \ref{Stimulation Position for Tasks}) can assist people with BLV in different tasks. 
For each task type, feedback method, and stimulation position, we discuss associated challenges and potential opportunities. 
Following these sections, we offer a detailed analysis of the limitations of existing haptic tools, focusing on three perspectives: 1) Hardware limitations (Section \ref{Hardware limitations}) including environmental effects, conspicuousness, and form factor; 
2) Functionality limitations (Section \ref{Functionality limitations}) including control over information display, information resolution, and presentation modality;
and 3) UX and evaluation method limitations (Section \ref{UX and evaluation method limitations}), including user diversity, customization, user study design, and user trust.
Figure \ref{fig:findings} shows the structure of these sections and subsections.

% Each paper was coded, and statistical calculation was conducted.

% We coded each paper based on x, y, z, and tallied the frequencies of each code to determine the distribution of x, y, z.

% Each paper was coded, and statistical calculation was conducted.
% Eighteen publications were excluded because they were posters. A further 13 papers were excluded because they did not use VR technology

% We screened the titles and abstracts of the 477 papers collected
% in Phase 1 by using the inclusion criteria presented above. The
% four authors individually rated the same set of 30 randomly chosen
% papers for inclusion. The overall percentage agreement on these 30
% papers was 91.11\%, and the Cohen’s Kappa 0.82, 95\% CI [0.68, 0.97].
% Two of the authors rated the rest of the papers for inclusion. Out of
% the 477, we included 80 papers for Phase 3 (of which 9 were in the
% set which was interrated), thereby excluding 397 papers (of which
% 21 were in the set which was interrated)

% The same set of 30 randomly chosen papers were individually rated by two authors for inclusion. Then, the three authors rated the rest of the papers for inclusion. 

\section{Haptic Assistive Tools for Various Tasks} \label{Haptic Assistive Tools for Various Tasks}
\begin{figure} [tbh!]
    \centering
    \includegraphics[scale=0.38]{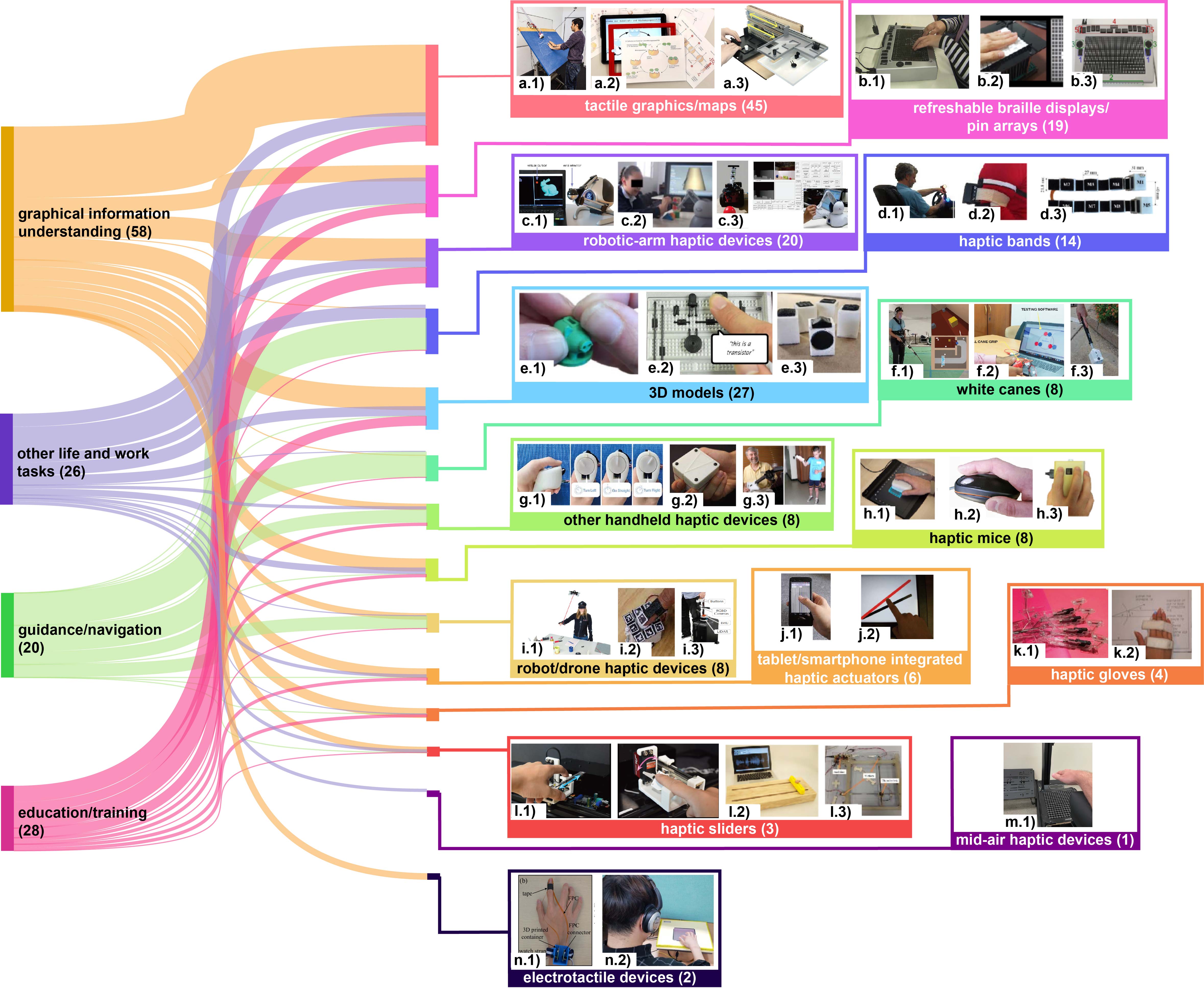}
     \caption{Examples of haptic assistive tools designed for the four main types of tasks. The connections show which haptic tools assist the specific BLV tasks.}
     \Description{This figure illustrates the examples of 14 haptic assistive tools designed for the four main types of tasks. The numbers after each haptic assistive tool is the number of paper utilized the device}
    \label{fig:task_haptic_assistive_tools}
\end{figure}

Based on the task that each haptic assistive tool was designed to support, we summarized them into four main types and tallied the number of papers that apply to each task. They include 1) graphical information understanding (58/132), 2) guidance/navigation (26/132), 3) education/training (20/132), and 4) other life and work tasks (28/132). We classified the haptic assistive tools for each task based on their form factor, which refers to their size, shape, and physical specifications. The form factors include common assistive tools (e.g., tactile graphics/maps, refreshable braille display/pin arrays) and other haptic devices that may not be frequently used as assistive tools (e.g., haptic mice, haptic gloves, and haptic bands). The classification of papers is shown in Table \ref{table:type_of_haptic_assistive_tool}, where the Sankey diagram (Figure \ref{fig:task_haptic_assistive_tools}) illustrates the distribution of haptic assistive tools that assist various BLV tasks.

\begin{table}[htbh!]
\caption{Type of Haptic Assistive Tool}
\centering 
\small
\begin{tabular}{p{3cm}p{2.5cm}p{2.5cm}p{2.5cm}p{2.5cm}}

\hline
\textbf{Type of Haptic Assistive Tool} & \textbf{Graphical information understanding} & \textbf{Guidance/navigation} & \textbf{Education/training} & \textbf{Other life and work tasks}\\\hline
tactile graphics/maps & \cite{swaminathan2016linespace,li2019editing,yang2020tactile,hofmann2022maptimizer,holloway20223D,Nagassa20233D,baker2014tactile,lucentmaps2016gotzelmann,reichinger2016gesture,suzuki2017fluxmarker,baker2014tactile,visually2018gotzelmann,reichinger2018pictures,gupta2019evaluating,zeinullin2022tactile,kim2015toward,braier2014haptic,mcGookin2010clutching,wang2009instant,ducasse2016tangible,roberts2005haptic,mascle2022tactile} & /  & \cite{albouys2018towards,melfi2020understanding,fusco2015tactile,holloway20193d,sargsyan20233D}  & \cite{das2023simphony,bornschein2015collaborative,shi2020molder,pandey2020explore} \\\hline
refreshable braille displays/pin arrays & \cite{prescher2017consistency,besse2018understanding,prescher2010tactile,rao20202across,Deschamps2012interpersonal} & / & \cite{petit2008refreshable,saikot2022refreshable,melfi2022audio} & \cite{bornschein2018comparing,Jung2021ThroughHand,siu2019shapecad,Morash2018evaluating,russomanno2015refreshing,Siu2019Advancing,Cassidy2013haptic,Pantera2021lotusbraille} \\\hline
tablet/smartphone integrated haptic actuators & \cite{palani2017principles,chu2022comparative} & / & \cite{Toennies2011toward} & \cite{Rantala2009methods} \\\hline
haptic mice & \cite{brayda2015the,kim2011handscope,levesque2012adaptive,pietrzak2009creating} & / & \cite{brayda2013predicting} & \cite{Headley2011roughness,Strachan2013vipong} \\\hline
haptic gloves & \cite{soviak2016tactile,quek2013enabling,chase2020pantoguide} & / & \cite{oliveira2012the} & / \\\hline
haptic sliders & \cite{fan2022slide,gay2021f2t} & / & / & \cite{Tanaka2016haptic}      \\\hline
robot/drone haptic devices  & \cite{guinness2019robo,ducasse2018botmap} & \cite{huppert2021guidecopter,guerreiro2019cabot,kayukawa2020guiding,rahman2023take,moon2019prediction} & \cite{guinness2019robo} & / \\\hline
haptic bands & / & \cite{hong2017evaluating,erp2020tactile,Kaul2021around,hirano2019synchronized,cosgun2014evaluation,lee2023novel,barontini2021integrating,flores2015vibrotactile,xu2020virtual,ogrinc2018sensory} & / & \cite{sucu2014the,Stearns2016evaluating,Allman2009rock,kayhan2022a}    \\\hline
3D models & \cite{shi2016tickers,reinders2020hey,brule2021beyond,holloway2023tacticons,shi2017designing,kane2013touchplates,shi2019designing} & / & \cite{holloway2018accessible,davis2020tangible,chang2021accessiblecircuits} & \cite{Guo2017facade,baldwin2017the,lee2023tacnote}    \\\hline
white canes & / & \cite{siu2020virtual,swaminathan2021from,nasser2020thermalcane,zhao2018enabling,tanabe2021identification,wang2012halo,kim2015identification} & / & /   \\\hline
robotic-arm haptic devices & \cite{abu_doush2009making,abu2010multimodal,moll2013haptic,de_felice2007haptic,sallnas2007group,schloerb2010blindaid,park2015telerobotic} & / & \cite{zhang2017multimodal,espinosa2021virtual,crossan2008multimodal,plimmer2011signing,murphy2015haptics} & \cite{lieb2020haptic,Plimmer2008multimodal,Park2013real}   \\\hline
other handheld haptic devices & \cite{yatani2012spacesense,sharmin2005first} & \cite{liu2021tactile,amemiya2009haptic,amemiya2010orienting,spiers2017design} & \cite{sanchez2010usability} & \cite{Morelli2010vi}   \\\hline
electrotactile devices & \cite{jiang2024designing,yoo2022perception} & / & / & / \\\hline

\end{tabular}
\label{table:type_of_haptic_assistive_tool} 
% \description{Table 1 describes}
\end{table}

\subsection{Haptic assistive tools for graphical information understanding}
\textit{Graphical information understanding} involves haptic assistive tools that support people with BLV with comprehending the overview and details of any figure, chart, or map \cite{jiang2023understanding}. \textbf{Tactile graphics/maps} are the most common assistive tools for this purpose (22/58), followed by \textbf{3D models} (7/58) and \textbf{robotic-arm haptic devices} (7/58). 
The \textbf{robotic-arm haptic devices} utilize a robotic arm, such as Phantom$^\circledR$, to guide the users with BLV during the learning process.
Its detailed applications for \textit{ducation/training} will be described in section \ref{Robotic_arm}.

\textbf{Tactile graphics/maps} are the images that use raised dots, lines, and surfaces to convey graphical information \cite{tg}, including maps, charts, and figures. 
In this context, maps depict the spatial layout or geographic relationships between various regions, charts are graphical tools used to represent numerical data in a structured way (e.g., bar or line charts), and figures are broader visual illustrations of objects or scenes.
The design considerations for different types of graphical information varied. 
In general, researchers focused on factors such as heights, widths, dashed patterns, and shapes of raised lines, dots, and surfaces to assist the three subtasks in graphical information understanding, including \textbf{map information understanding}, \textbf{chart information understanding}, and \textbf{figure information understanding}.

For \textbf{map information understanding}, the researchers often used a single tactile graphic to display the whole map.
They first determined which geographic features to include---such as roads, buildings, and other facilities \cite{hofmann2022maptimizer,Nagassa20233D,gupta2019evaluating}. 
They then decided how to represent these selected features. 
Some used raised lines to indicate roads, varying the width and dashed patterns to represent different road types, from footpaths to highways \cite{hofmann2022maptimizer}. 
Others employed raised spaces and raised dots of various shapes (e.g., triangles, squares) to convey textures representing different map areas \cite{hofmann2022maptimizer,holloway20223D,gupta2019evaluating}. 
Additionally, some researchers utilized movable 3D-printed icons to denote specific landmarks, such as restrooms or shops \cite{Nagassa20233D}.

\textbf{Chart information understanding} followed a similar approach, with the primary distinction being that researchers focused on which chart features to include rather than geographic features. 
This included methods to assist users with BLV in distinguishing various bars, dots, lines, and slices \cite{braier2014haptic,baker2014tactile}.

For \textbf{figure information understanding}, researchers typically began by separating figures into various areas and layers based on their structures. 
For example, they separated the elephant's figure into its leg, body, ear, head, and teeth according to its taxonomy structure \cite{kim2015toward,gupta2019evaluating,reichinger2018pictures}. 
These different areas were generally displayed with height variations that reflected their real-world perspectives \cite{kim2015toward,gupta2019evaluating}. 
In contrast to maps and charts, which can be simplified into key elements (e.g., geographic relationships or bar heights), figures are typically represented as closely as possible to the original visualization \cite{reichinger2018pictures}.
% In contrast to map information, figure information was usually represented similarly to the original visualized figure \cite{reichinger2018pictures}. 
Additionally, while map information often used a single tactile graphic to represent the entire map, figure information sometimes employed multiple tactile graphics to convey different elements.

\begin{figure} [tbh!]
    \centering
    \includegraphics[scale = 0.8]{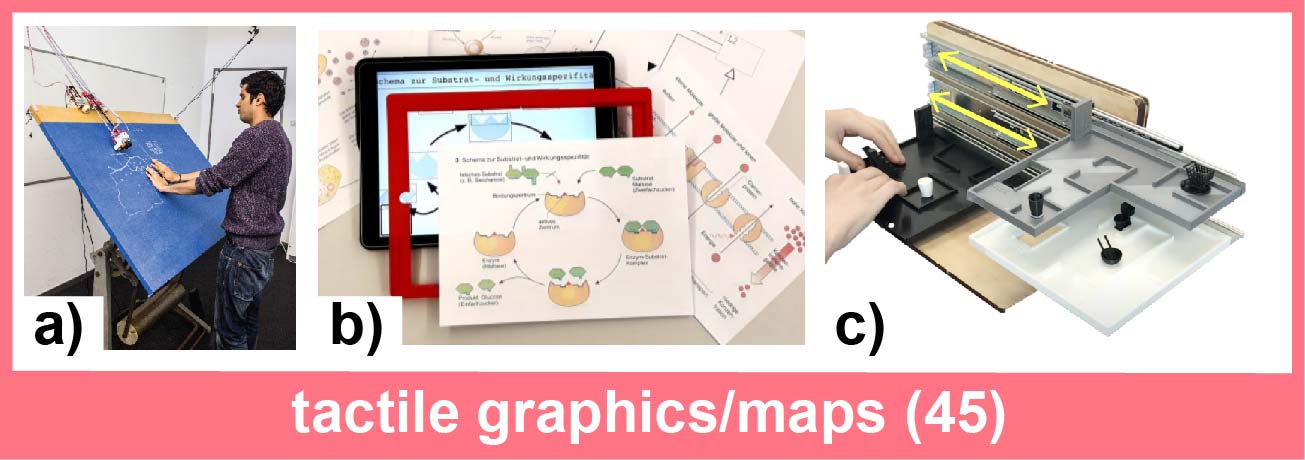}
     \caption{Examples of tactile graphics/maps: a) Linespace: a refreshable tactile graphic system based on 3D printers \cite{swaminathan2016linespace}; b) a mobile audio-tactile system for learning graphics at schools \cite{melfi2020understanding}; c) 3D Building Plans: a 3D-printed multi-story building tactile map for O\&M education. The figure shows one of the three designs in which the users can slide away each layer of the tactile map \cite{Nagassa20233D}.}
    \Description{This figure illustrates the examples of tactile graphics/maps, each figure includes three subfigures. Each subfigure shows one prototype.}

    \label{fig:tactile_graphics}
\end{figure}

The design differences between maps and figures were informed by the nature of the information being conveyed to users with BLV.
Map information design focused more on providing an overview, often omitting detailed elements or relying on audio feedback for additional details. 
This is because tactile graphics offer lower spatial acuity compared to visual perception, as obtaining information through touch is inherently more “sequential” than through vision. 
In contrast, audio is not limited by spatial constraints and instead conveys information temporally \cite{morash2012review}.
In comparison, the specifics were crucial for figure information, necessitating designs that closely resembled the original figures.
An example of maps is a 3D Building Plan developed by Nagassa et al. to show multi-story buildings' indoor structures. They found that users with BLV preferred the overlapped presentations, a movable stack of each floor's tactile map, shown in Figure \ref{fig:tactile_graphics} c). The overlapped presentations enabled more effective building of cross-floor spatial overview \cite{Nagassa20233D}. Swaminathan et al. leveraged 3D printers to create a refreshable tactile display system by printing the raised lines of filament in real-time, shown in Figure \ref{fig:tactile_graphics} a). They found that their design is more expressive compared to the traditional refreshable braille displays which present graphics with dots \cite{swaminathan2016linespace}. 

\textbf{3D models} are the second most used haptic assistive tools to assist in graphical information understanding, including either wooden or 3D-printed models that convey spatial information \cite{holloway2023tacticons}.

When designing these models, researchers focused on ensuring they could be easily distinguished by users. 
Some created 3D models based on real-world objects and referenced familiar visual icons, such as flowers, signs, and buildings, which are easily recognizable and memorable \cite{holloway2023tacticons,shi2017designing,shi2019designing}. 
They transformed these 2D icons into 3D models by following their real-world outlines \cite{holloway2023tacticons}. 
Additionally, some researchers incorporated removable components to allow users to explore the details. 
These models closely resembled actual structures and were often utilized in formal contexts, such as science popularization and education/training \cite{reinders2020hey}. 
Recently, 3D models have commonly been integrated with audio labels, supporting users' understanding of various parts of the models, such as continents on a globe or components within a cell \cite{shi2019designing,shi2016tickers}.

\begin{figure} [tbh!]
    \centering
    \includegraphics[scale = 0.8]{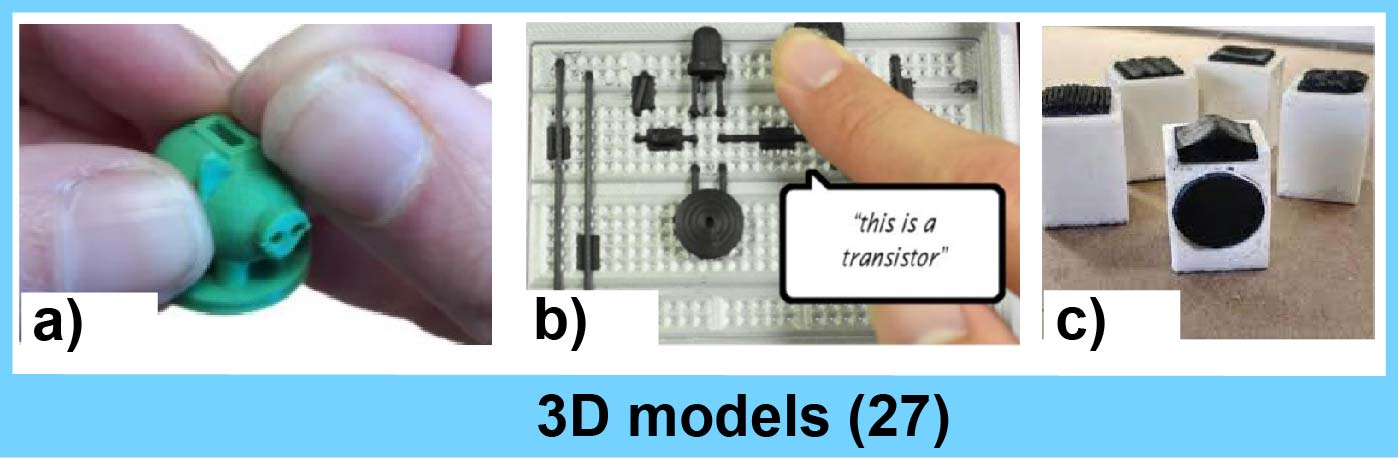}
     \caption{Examples of 3D models: a) TactIcons: 3D printed icons that can substitute visual icons for street and park tactile maps \cite{holloway2023tacticons}; b) TangibleCircuits: a haptic and audio feedback device based on 3D models that allow users with BLV to understand circuit diagrams \cite{davis2020tangible}; c) Tangible Desktop: 3D-printed icons for multimodal interaction of computers. Each icon has an RFID tag embedded inside and a tactilely distinct rubber crown \cite{baldwin2017the}.}
     \Description{This figure illustrates the examples of 3D models, each figure includes three subfigures. Each subfigure shows one prototype.}
    \label{fig:3D_models}
\end{figure}

For example, Holloway et al. developed a series of 3D-printed icons to assist people with BLV in recognizing landmarks on maps as efficiently as sighted people, which is shown in Figure \ref{fig:3D_models} a). They found that participants with BLV can learn their 3D-printed icons easily and recognize them instantly. \cite{holloway2023tacticons}. Davis et al. presented a multimodal device that allows users with BLV to understand circuit diagrams by interacting with 3D-printed circuit tangible models, which is shown in Figure \ref{fig:3D_models} b). They found that users with BLV can better understand the geometric, spatial, and structural circuit information using their device \cite{davis2020tangible}.

\begin{figure} [tbh!]
    \centering
    \includegraphics[scale = 0.8]{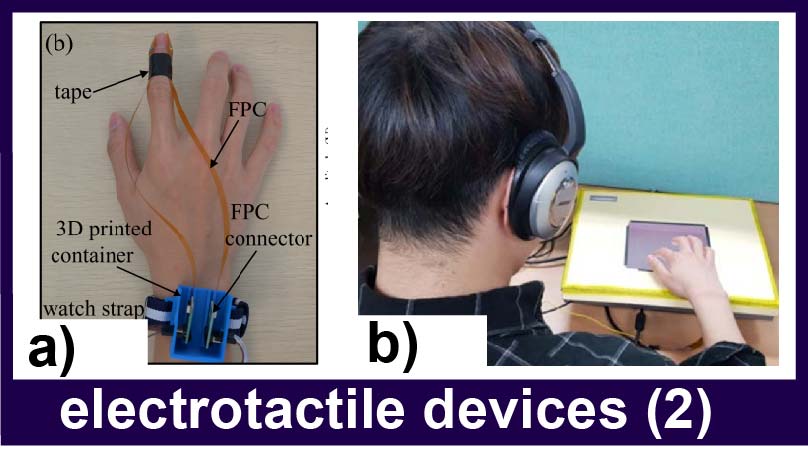}
     \caption{Examples of Electrotactile Devices: a) an assistive tool based on unobtrusive modulated electrotactile feedback on fingertip Edge for charts comprehension \cite{jiang2024designing}; b) An electrostatic skin-stretch display for graphics exploration  \cite{yoo2022perception}.}
     \Description{This figure illustrates the examples of electrotactile devices with two subfigures. Each subfigure shows one prototype.}
    \label{fig:electrotactile_devices}
\end{figure}

Additionally, electrotactile devices, illustrated in Figure \ref{fig:electrotactile_devices}, utilize electric current applied directly to the skin or through a thin insulating layer with conductive material to generate tactile sensations \cite{yoo2022perception,jiang2024designing}. 
Researchers have focused on various features of modulated electrical signals, including voltage, current, frequency, stimulation duration, and intervals \cite{lin2022super}.

\subsection{Challenges in graphical information understanding}
Based on our literature survey, we synthesized the main challenges that exist in \textit{graphical information understanding}, which include \textbf{low comprehension efficiency of the overview}, \textbf{difficulty with detail gathering}, and \textbf{high short-term memory load} \cite{fan2023accessibility, baldwin2017the}. 
The challenges of \textbf{low comprehension efficiency of the overview} and \textbf{difficulty with detail gathering} arise because individuals with BLV must rely on audio descriptions of alternative text or explore tactile graphics with their fingers to understand details at each location. 
This step-by-step process to construct an overview is more time-consuming than the experience of sighted individuals \cite{jiang2023understanding}. 
Moreover, building an overview requires people with BLV to retain details quickly and precisely, which imposes a \textbf{high short-term memory load} \cite{baldwin2017the}.

\textbf{Low comprehension efficiency of the overview} and \textbf{difficulty with detail gathering}: 
Haptic devices often support users in grasping both the overview and intricate details of figures or maps. 
Initially, these devices guide users through the overall structure to establish a foundational understanding of the charts or figures. 
Subsequently, they facilitate free exploration to deepen comprehension. For instance, Abu-doush's robotic-arm haptic devices offer guided and exploratory modalities.
The guided mode directs users' fingers along predefined trajectories to comprehend trends in line charts, while the exploratory mode enables users to explore individual bars or slices in bar or pie charts \cite{abu2010multimodal}.

\textbf{High short-term memory load}: 
High short-term memory load presents another obstacle with graphical information understanding. 
Previous studies identified note-taking as a strategy to alleviate this burden \cite{lee2023tacnote, pandey2020explore}.
However, for individuals with BLV, conventional note-taking methods can be challenging due to the difficulty in retrieving data and the inconvenience of switching between note-taking applications \cite{jiang2023understanding}.
To address this issue, Lee et al. investigated the use of a 3D printing pen to facilitate note-taking by creating tangible 3D models.
Through tactile exploration, users with BLV could specify the locations, ordering, and hierarchy relationships of notes, effectively reducing their memory load \cite{lee2023tacnote}.
Despite the potential of haptic devices to address various challenges, most focus solely on aiding comprehension of overviews and details. 
Few explore methods to mitigate short-term memory load, highlighting a promising area for further research.

\subsection{Haptic assistive tools for guidance/navigation}
\textit{Guidance/Navigation} involves haptic assistive tools that support people with BLV with finding directions or specific objects in indoor and outdoor scenarios \cite{csapo2015survey}. \textbf{Haptic bands} (10/26) and \textbf{white canes} (8/26) are the most used haptic assistive tools. 

The \textbf{haptic bands} are bands that are equipped with haptic actuators (e.g., vibrators) and are bound to the user's arm, wrist, or waist. 
The design of haptic bands typically focused on accurately conveying guidance cues and mapping them to real-world directions. 
Researchers first considered the arrangement of actuators around the bands, including their number, type, and density \cite{hong2017evaluating}. 
Normally, the higher number of vibrators could ensure denser arrangments, and thus convey more precise guidance cues \cite{hong2017evaluating}.
They then determined the vibration patterns for guidance and notification cues, which included both single and sequential vibrations \cite{hong2017evaluating,xu2020virtual}. 
Single vibrations activated the vibrator closest to the target direction \cite{flores2015vibrotactile}, while sequential vibrations involved more complex patterns, with the direction indicated from the vibration start point to the endpoint \cite{hong2017evaluating, Kaul2021around, ogrinc2018sensory,lee2023novel}. 
Some designs incorporated multiple vibrations, where rotational vibrations around the arms or wrists signified turning directions—clockwise for right turns and counterclockwise for left turns \cite{ogrinc2018sensory}.

Additionally, vibration durations, intervals, and overlaps were considered to convey notifications. 
The combination of different stimulation durations and pauses can indicate various states, such as attention, walking, stopping, and barriers ahead of the users \cite{lee2023novel}.

\begin{figure} [tbh!]
    \centering
    \includegraphics[scale = 0.8]{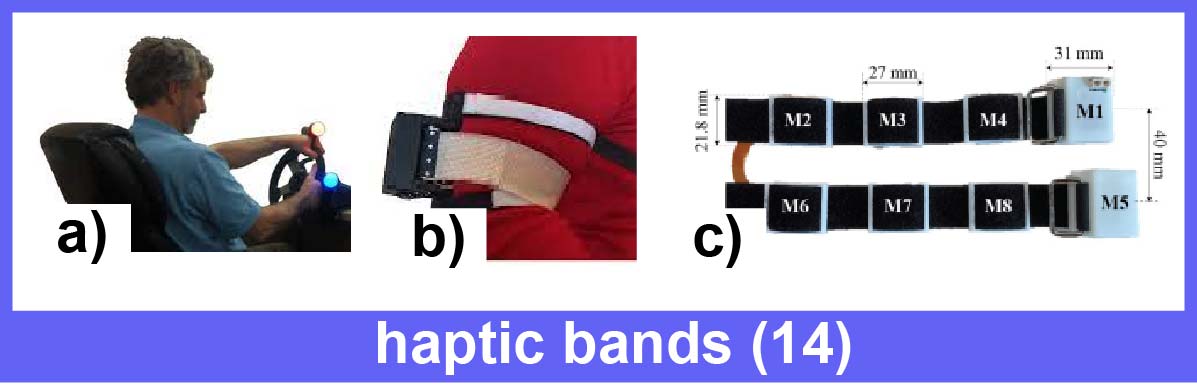}
     \caption{Examples of haptic bands: a) A vibrotactile band worn on the hand to assist drivers with BLV through vibration cues \cite{sucu2014the}; b) A haptic band based on vibration worn on the users with BLV's arms to provide guidance cues \cite{barontini2021integrating}; c) A haptic band worn on the arm to assist users with BLV in independent travel through eight vibrators \cite{lee2023novel}.}
     \Description{This figure illustrates the examples of haptic bands, each figure includes three subfigures. Each subfigure shows one prototype.}
    \label{fig:haptic_bands}
\end{figure}

For example, Lee et al. presented a wrist-worn vibrotactile device to provide directional, spatial, and status cues to assist people with BLV's independent walking and wayfinding, shown in Figure \ref{fig:haptic_bands} c). They found that their vibrotactile information was quickly identified and it enabled users with BLV to walk independently without any pre-learned knowledge about the route\cite{lee2023novel}. Barontini et al. developed a wearable system to assist people with BLV in indoor navigation and obstacle avoidance, shown in Figure \ref{fig:haptic_bands} b). They found that their device could support indoor navigation and allowed people with BLV to travel independently \cite{barontini2021integrating}. 

\textbf{White canes} are the second most used haptic assistive tools to assist in guidance/navigation. White canes are sticks between 100 to 150 centimeters in length that help people with BLV detect obstacles on the road, which may be equipped with either thermal modules \cite{nasser2020thermalcane} or vibrators \cite{siu2020virtual}.

The design of white canes focused on effectively conveying directional cues and notifying users of barriers, including their direction, distance, and height \cite{wang2012halo,kim2015identification,siu2020virtual}. 
Notification cues could be delivered through vibrators and mechanical brakes \cite{siu2020virtual}. 
Some researchers utilized brakes to provide resistance and vibrators to simulate various textures encountered on the ground, such as carpets, tactile pavements, or walls \cite{siu2020virtual,zhao2018enabling,tanabe2021identification}. 
Obstacles were typically indicated by single vibrations to convey their locations (using different positions on the white cane) and heights (through varying vibration frequencies) \cite{wang2012halo,kim2015identification}.

Additionally, some studies explored how to convey directional cues using white canes, with design considerations similar to those of haptic bands, including the number, type, and density of actuators. 
The key difference was that white canes were held by users rather than worn on their arms or wrists, meaning that the holding posture affected how guidance cues were mapped, which was an important consideration in their design \cite{nasser2020thermalcane}.

\begin{figure} [tbh!]
    \centering
    \includegraphics[scale = 0.8]{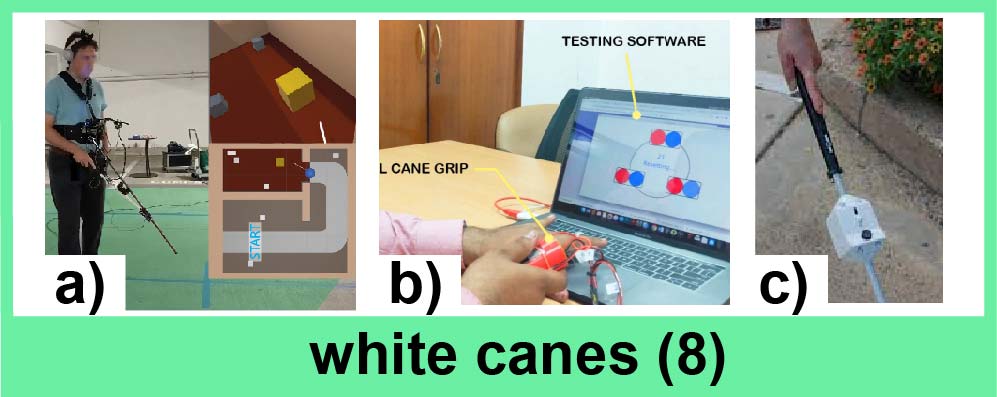}
     \caption{Examples of white canes: a) A white cane integrated with vibrators to guide users with BLV in VR environment \cite{siu2020virtual}; b) A white cane integrated with Peltiers that can guide users with BLV via thermal feedback \cite{nasser2020thermalcane}; c) HALO: A white cane integrated with vibrators to assist people with BLV in recognizing low-hanging obstacles such as tree branches and open cabinet doors \cite{wang2012halo}.}
     \Description{This figure illustrates the examples of white canes, each figure includes three subfigures. Each subfigure shows one prototype.}
    \label{fig:white_canes}
\end{figure}

For example, Nasser et al. developed Thermalcane, a white cane that can offer thermotactile directional cues for users with BLV, shown in Figure \ref{fig:white_canes} b). They found that the thermal feedback yielded significantly higher accuracy than the vibrotactile feedback used with white cane \cite{nasser2020thermalcane}. Siu et al. developed a multimodal cane controller that provides texture simulation and spatialized audio, shown in Figure \ref{fig:white_canes} a). They found that their device allows people with BLV to navigate in large virtual environments with complex architecture without vision \cite{siu2020virtual}.

\subsection{Challenges in guidance/navigation}
The main challenges of \textit{guidance/navigation} based on our literature review include \textbf{disorientation with directions}, \textbf{difficulty with avoiding obstacles and crowds}, \textbf{risk of falling}, and \textbf{risks associated with intersection crossings} \cite{riazi2016outdoor}. 

The challenge of \textbf{disorientation with directions} arises because individuals with BLV often struggle to identify landmarks, making it difficult to understand their current location and decide on the next steps. 
Additionally, the lack of visual input complicates the recognition of approaching pedestrians, obstacles, and height differences between road surfaces. 
This, in turn, leads to challenges such as \textbf{difficulty avoiding obstacles and crowds}, an increased \textbf{risk of falling}, and heightened \textbf{risks associated with crossing intersections} \cite{riazi2016outdoor}.

\textbf{Disorientation with directions}: 
Certain haptic devices address the challenge of disorientation with directions via haptic cues. 
For instance, tactile graphics/maps, mid-air haptic devices, and haptic sliders offer tactile representations of maps, enabling users with BLV to familiarize themselves with environments prior to actual navigation \cite{Nagassa20233D,fink2023autonomous,gay2021f2t}.

\textbf{Difficulty with avoiding obstacles and crowds}: 
Haptic notifications are provided by tools like haptic bands, white canes, handheld haptic devices, and robot/drone haptic devices, aiding users in navigation and obstacle avoidance  \cite{lee2023novel,barontini2021integrating,liu2021tactile,huppert2021guidecopter,nasser2020thermalcane,kayukawa2020guiding,guerreiro2019cabot}.
These devices offer real-time cues for independent travel, alerting users to obstacles or pedestrians along their path.

\textbf{Risk of falling}: 
The risk of falling is effectively addressed by white canes and robots, as they offer physical support between users with BLV and the ground, substantially reducing the likelihood of falls \cite{nasser2020thermalcane,kayukawa2020guiding,guerreiro2019cabot}. 
Unlike haptic bands, handheld haptic devices, and drones, which are either worn or held by users, white canes and guiding robots can establish robust physical support, enhancing users' stability and safety during navigation \cite{lee2023novel,barontini2021integrating,liu2021tactile,huppert2021guidecopter}. 

\textbf{Risks associated with intersection crossings}: 
Our literature review did not identify any papers specifically addressing the challenge of \textit{risks associated with intersection crossings}. 
While previous studies have investigated auditory feedback to aid individuals with BLV in street crossing, concerns about auditory distractions from surrounding noise have been raised \cite{borenstein1997guidecane,velazquez2010wearable}. 
Therefore, there is potential for future research to explore the use of haptic cues to assist individuals with BLV in safely navigating intersections.
Developing haptic devices capable of providing guidance cues and preventing falls could greatly enhance the safety and independence of individuals with BLV during street crossings.

\subsection{Haptic assistive tools for education/training}
% \textit{Education/Training} is how people with BLV are assisted in STEM education, Orientation and Mobility (O\&M) training, and other skill learning. 
\textit{Education/Training} includes various disciplines such as STEM education, Orientation and Mobility (O\&M) training, and other skill learning. 
people with BLV participate O\&M training to learn skills to travel safely and efficiently through their environment, such as wayfinding/orientation, street crossings, public transit, and use of GPS and electronic travel aids \cite{wiener2010foundations}. Papers on assistive tools for O\&M training study how the devices can be used to train users in a controlled environment with instructors, which is different from guidance/navigation tasks in which the users travel alone with the tools' assistance. \textbf{Tactile graphics/maps} (5/20) and \textbf{robotic-arm haptic devices} (5/20) are the most used haptic assistive tools for this task. 

For \textbf{tactile graphics/maps}, design considerations are similar to those for \textbf{graphical information understanding}. 
A key consideration is that researchers often designed tactile graphics/maps to be detachable or include movable 3D icons, allowing instructors to let students with BLV explore them separately, which facilitates progressive teaching \cite{sargsyan20233D,holloway20193d}. 
Additionally, enabling teachers to easily modify audio descriptions for various locations and facilities on the tactile graphics/maps is another important design aspect. 
This flexibility allowed educators to guide students from basic recognition to a deeper understanding, transitioning from simply knowing names to comprehending the specific functions of different locations and facilities \cite{sargsyan20233D,fusco2015tactile,melfi2020understanding}.

For example, Melfi et al. introduced a mobile audio-tactile learning environment to facilitate students with BLV's visual information learning, shown in Figure \ref{fig:tactile_graphics} b). They found that their device is suitable for educational environments because it allows faster exploration of tactile graphics and better memorization of information \cite{melfi2020understanding}. Sargsyan et al. developed 3D-printed tactile maps of a train station to teach the people with BLV about complex spatial knowledge, part of  O\&M training. They found that complex spatial notions are better understood with their 3D maps compared to traditional 2D tactile maps \cite{sargsyan20233D}. 

\textbf{Robotic-arm haptic devices} are the second most used haptic assistive tools to assist in education/training. 
\label{Robotic_arm}
% The \textbf{robotic-arm haptic devices} utilize a robotic arm, such as Phantom $^\circledR$, to guide the users with BLV during the learning process, including two subtasks: \textbf{2D structures learning} and \textbf{3D structures learning.}
The \textbf{robotic-arm haptic devices} were utilized for two subtasks: \textbf{2D structures learning} and \textbf{3D structures learning.}

Due to their high degree of freedom (DOF), the design of robotic-arm haptic devices emphasized encoding 2D and 3D information through the cursor's movement, force, and vibration. 

For \textbf{2D structures learning}, the robotic arm typically operated in an x-y plane, parallel to the ground, with movements indicating the outlines of figures or the trends of charts \cite{zhang2017multimodal,crossan2008multimodal,plimmer2011signing}. 
Applied forces, such as gravity acting vertically and viscosity opposing the user’s movement, helped notify users when they are on the correct trajectory along a figure's outline or a line chart \cite{zhang2017multimodal,abu2010multimodal}. 
This design rationale ensured that users perceive certain areas as more significant, encouraging them to continue moving along the designated lines or locations \cite{zhang2017multimodal,abu2010multimodal}. 
Additionally, vibrators installed on the cursor convey notification cues, indicating whether users are out of bounds or on a specific data point, thereby signaling when they are in a special location on the figure or chart \cite{abu2010multimodal}.

For \textbf{3D structures learning}, design considerations are similar to those of 2D figures, but the cursor moves within a 3D space. 
Researchers often employed forces to signal collisions with 3D objects and use vibrations to mark points of interest \cite{murphy2015haptics,espinosa2021virtual}.

In addition to facilitating independent learning, researchers also focused on increasing teacher involvement in the educational process. 
One design allows the robotic arm to follow the teacher’s movements in real-time, enabling students to mimic these movements instead of the pre-defined trajectories. 
This approach empowers teachers to instruct and correct students' mistakes immediately \cite{plimmer2011signing}.

% For example, Espinosa et al. explored how haptic feedback can support students with BLV in learning math, particularly fundamental three-dimensional (3D) shapes, shown in Figure \ref{fig:robotic_arm_haptic_devices} b). They found that haptic virtual perception is a valid and effective assistive technology for the education of blind children \cite{espinosa2021virtual}. Zhang et al. developed a real-time multimodal image perception system with the user of robotic arms to assist students with BLV'STEM course learning. They found that their device produced higher accuracy in recognizing images compared to conventional tactile images \cite{zhang2017multimodal}. 

\begin{figure} [tbh!]
    \centering
    \includegraphics[scale = 0.8]{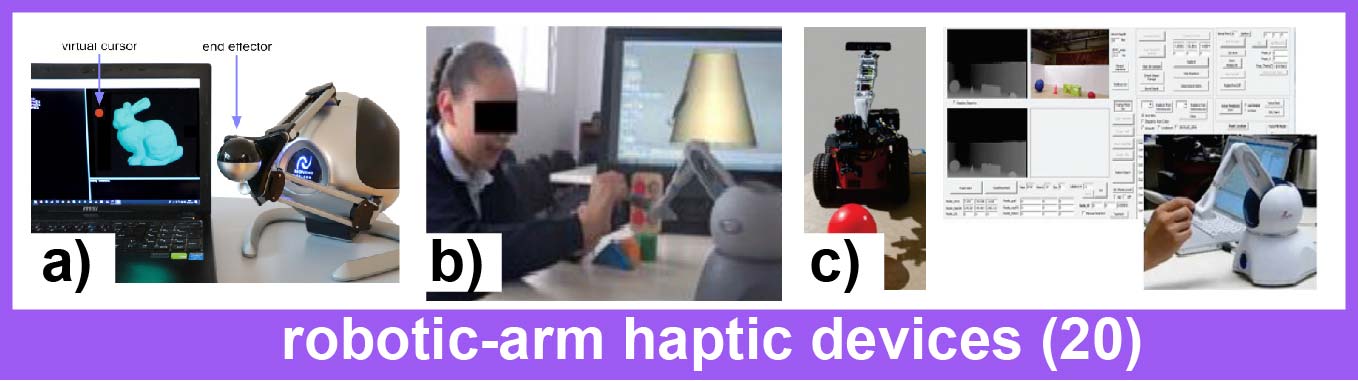}
     \caption{Examples of robotic-arm haptic devices: a) a multimodal robotic-arm haptic device that assists people with BLV to build 3D models independently \cite{lieb2020haptic}; b) A robotic-arm haptic device that assists people with BLV's education. The figure shows how it supports the students with BLV in comprehending various shapes \cite{espinosa2021virtual}; c) A robot and robotic arm system to assist people with BLV in telepresence \cite{Park2013real}.}
     \Description{This figure illustrates the examples of robotic-arm haptic devices, each figure includes three subfigures. Each subfigure shows one prototype.}
    \label{fig:robotic_arm_haptic_devices}
\end{figure}

\subsection{Challenges in education/training}
For \textit{education/training}, challenges include \textbf{difficulty following the blackboard and slice presentations}, \textbf{difficulty communicating with sighted classmates}, and \textbf{low efficiency in learning knowledge based on graphical information}. 

The challenge of \textbf{difficulty following the blackboard and slice presentations} is induced when people with BLV are unable to identify the specific location where the teacher is pointing—whether with a hand during face-to-face classes or with a cursor in online settings. 
Additionally, the lack of visual access contributes to \textbf{low efficiency in learning knowledge based on graphical information}, mirroring the challenges associated with \textit{graphical information understanding}. 
This divergence in learning processes, coupled with an inability to fully follow the teacher’s explanations, often leads to \textbf{difficulty following the blackboard and slice presentations}.

\textbf{Difficulty following the blackboard and slice presentations}: 
Haptic gloves have been utilized to address the challenge of difficulty following the blackboard and slice presentations. 
Oliveira et al. devised a system that tracked both the instructors' hand movements on the blackboard and the students with BLV's hand movements on tactile graphics, which were proportionally scaled representations of the graphics displayed on the blackboard.  
Vibration feedback guided the students' hands to the corresponding positions on the tactile graphics based on where the instructors' hands pointed on the blackboard \cite{oliveira2012the}. 
However, the system's bulkiness and large size raise concerns about its impact on the learning experience of sighted classmates, potentially obstructing their view of the presentations. 
Hence, there is a need to address how students with BLV can effectively follow instructors without impeding the learning of other students.

\textbf{Difficulty communicating with sighted classmates}: 
No paper in our literature review looked into how to support students with BLV in effectively communicating their ideas and knowledge with sighted classmates.
Researchers have highlighted that the lack of communication of class content with classmates could potentially diminish students' learning efficiency \cite{you2021difference}. 
Although it is important for students with BLV to effectively communicate with their sighted peers, they typically rely on haptic assistive tools, such as tactile graphics, to convey precise graphical information by indicating specific lines and areas \cite{d2021accessible}. 
Future research could focus on facilitating students with BLV in creating their own tactile graphics more efficiently, which could enhance their ability to communicate with sighted classmates.

\textbf{Low efficiency in learning knowledge based on graphical information}:  
Tactile graphics/maps, refreshable braille displays, 3D models, and robotic-arm haptic assistive tools have emerged as effective solutions to address the challenge of low efficiency in learning knowledge based on graphical information \cite{albouys2018towards,holloway2018accessible,saikot2022refreshable,zhang2017multimodal,espinosa2021virtual}. 
These tools empower students with BLV to explore figures, charts, and maps by utilizing finger movement and sensing height differences, enabling them to comprehend the spatial relationships within these graphical representations  \cite{albouys2018towards,holloway2018accessible,saikot2022refreshable}. Additionally, robotic-arm haptic assistive tools offer both guided navigation and free exploration, facilitating users in developing an overview through guidance and comprehending details through exploration \cite{zhang2017multimodal,espinosa2021virtual}. 
However, existing haptic assistive tools primarily target commonly used charts such as line charts, scatterplots, bar charts, and pie charts, with limited exploration of more complex chart types like radar charts, heat maps, and Sankey diagrams  \cite{fan2022slide,abu2010multimodal,chase2020pantoguide}.

In the short term, researchers have an opportunity to develop haptic devices capable of supporting students with BLV in comprehending these less-studied chart formats.
However, these formats were designed to align with visual perceptual patterns, which may not be optimal for non-visual modalities. 
In the long term, researchers could explore more effective non-visual approaches to data sensemaking, focusing on tactile-first and audio-first perspectives. These advancements could lead to a broader set of guiding principles for non-visual representation, allowing haptic devices to better support diverse data comprehension needs.
% Hence, there is a short-term opportunity for researchers to innovate and develop haptic devices capable of supporting students with BLV in comprehending these less-studied chart formats.
% In the long term, researchers could investigate more effective non-visual approaches for data sensemaking.
% Based on these advancements, haptic devices leveraging the new methodologies could be developed to enhance data acquisition for users with BLV.

\subsection{Haptic assistive tools for other life and work tasks}
\textit{Other life and work tasks} consisted of the subtasks that people with BLV conduct during their daily life and work, which less than five papers focused on. 

These subtasks include hand drawing/writing \cite{bornschein2018comparing,Plimmer2008multimodal,pandey2020explore}, braille reading \cite{Morash2018evaluating,russomanno2015refreshing,Rantala2009methods,Pantera2021lotusbraille}, gaming \cite{Jung2021ThroughHand,Allman2009rock,Morelli2010vi,Strachan2013vipong}, interacting with appliances \cite{Guo2017facade,baldwin2017the,Cassidy2013haptic}, using 3D modeling software \cite{siu2019shapecad,lieb2020haptic,Siu2019Advancing}, driving \cite{fink2023autonomous,sucu2014the}, designing assistive tools \cite{bornschein2015collaborative,shi2020molder}, presenting the roughness texture on tactile graphics \cite{Headley2011roughness}, weaving \cite{das2023simphony}, taking notes \cite{lee2023tacnote}, producing audio work \cite{Tanaka2016haptic}, and providing telepresence \cite{Park2013real}. 
\textbf{Refreshable braille displays/pin arrays} (8/28), \textbf{tactile graphics/maps} (4/28), and \textbf{haptic bands} (4/28) are the most used haptic assistive tools.

\begin{figure} [tbh!]
    \centering
    \includegraphics[scale = 0.8]{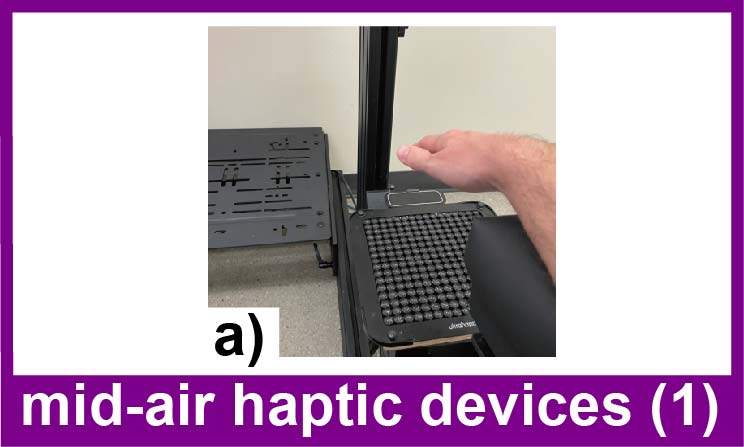}
     \caption{Examples of mid-air devices: A mid-air haptic device to assist drivers with BLV in comprehending street maps \cite{fink2023autonomous}.}
    \label{fig:mid_air_devices}
    \Description{This figure illustrates the examples of mid air devices, each figure includes three subfigures. Each subfigure shows one prototype.}
\end{figure}

\begin{figure} [tbh!]
    \centering
    \includegraphics[scale = 0.8]{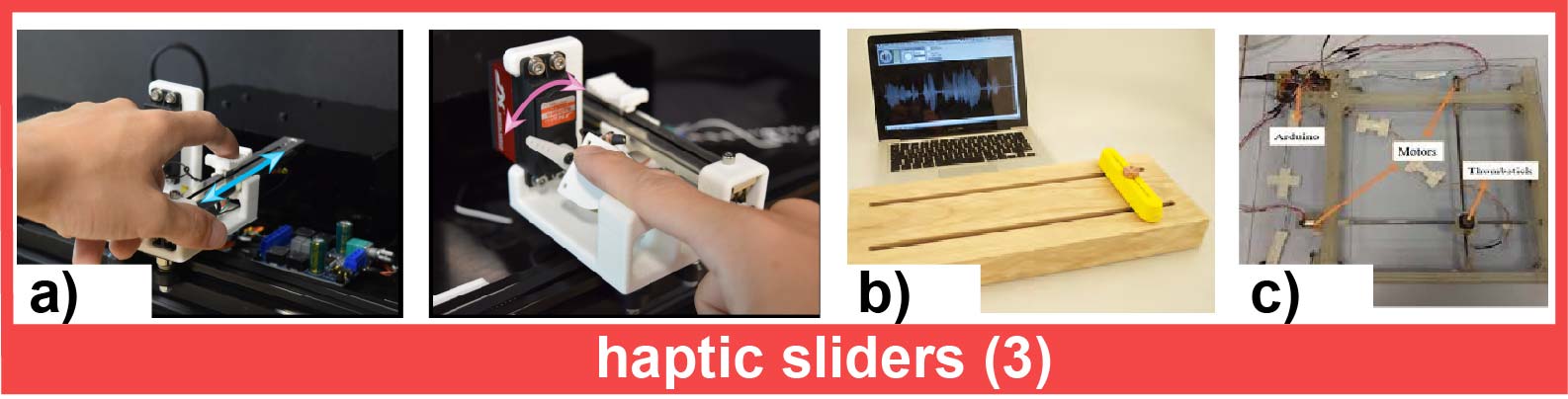}
     \caption{Examples of haptic sliders: a) Slide-Tone and Tilt-Tone: a system that assists users with BLV in comprehending line charts. The figures showed the designs of the Slide-Tone device which conveys y-axis value in each x-axis value, and the Tilt-Tone device which conveys inclination in each x-axis value \cite{fan2022slide}; b) Haptic Wave: a haptic slider to assist people with BLV in audio producing \cite{Tanaka2016haptic}; c) F2T: a haptic slider device that assists people with BLV in understanding 2D maps \cite{gay2021f2t}.}
     \Description{This figure illustrates the examples of haptic sliders, each figure includes three subfigures. Each subfigure shows one prototype.}
    \label{fig:haptic_sliders}
\end{figure}

\textbf{Refreshable braille display/pin array} is an electro-mechanical device for displaying braille characters and figures by round-tipped pins and surfaces \cite{rbd}. For example, Siu et al. developed a 2.5D shape display that enables designers with BLV  to explore and modify existing models\cite{siu2019shapecad}. Jung et al. developed a device that enables users with BLV to interact with multiple dynamic objects in real-time and thus play 2D games \cite{Jung2021ThroughHand}. 

For \textbf{tactile graphics/maps}, Das et al. presented an audio-tactile system that aims to support blind weavers in creating and perceiving patterns. They revealed how blind weavers used their devices to learn the process of pattern design and generate patterns with sighted instructors \cite{das2023simphony}. Lee et al. explored opportunities to enable object-centric note-taking using a 3D printing pen for interactive, personalized tactile annotations. Their device efficiently assisted users with BLV in notetaking \cite{lee2023tacnote}. 

\begin{figure} [tbh!]
    \centering
    \includegraphics[scale = 0.8]{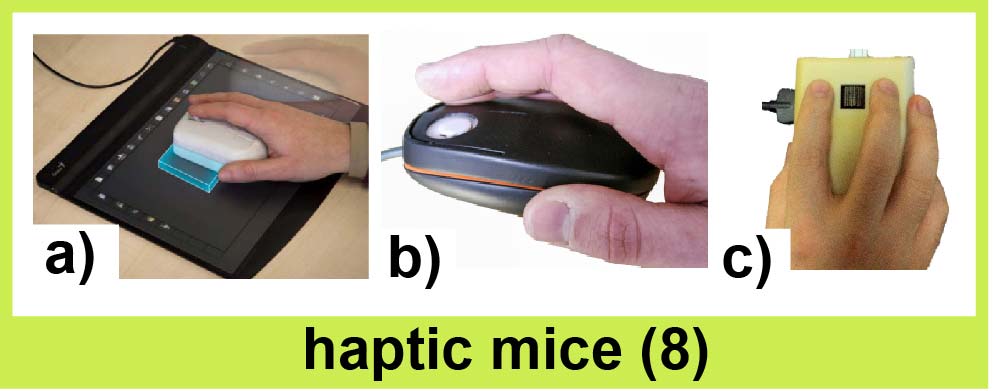}
     \caption{Examples of haptic mice: a) a haptic mouse that assists people with BLV in comprehending 3D shapes for their O\&M education \cite{brayda2013predicting}; b) ViPong: a haptic mouse to assist people with BLV in playing games \cite{Strachan2013vipong}; c) a haptic mouse that assists people with BLV in understanding graphics with different details \cite{Park2013real}.}
     \Description{This figure illustrates the examples of haptic mice, each figure includes three subfigures. Each subfigure shows one prototype.}
    \label{fig:haptic_mice}
\end{figure}

Additionally, researchers also presented other haptic assistive tools to support interacting with appliances via 3D models \cite{Guo2017facade,baldwin2017the}, driving via mid-air haptic devices \cite{fink2023autonomous} shown in Figure \ref{fig:mid_air_devices} a),  and haptic bands \cite{sucu2014the}, assisting audio production with haptic sliders shown in Figure \ref{fig:haptic_sliders} b) \cite{Tanaka2016haptic}, providing roughness on graphics with haptic mice\cite{Headley2011roughness} shown in Figure \ref{fig:haptic_mice} c), and providing telepresence via robotic-arm haptic devices \cite{Park2013real}.
Mid-air haptic devices could deliver unobtrusive haptic feedback directly to users' hands without requiring physical contact, enabling remote interactions. 
Although only one paper in our corpus explored the application of a mid-air assistive tool, its advantages---such as ease of use and remote interaction—suggest its potential for wider application in public assistive tools \cite{paneva2020haptiread}.

% The mid-air haptic devices could deliver unobtrusive haptic information directly to the user without touching any device, which allowed remote interactions.
% Though in our venue only one paper introduced the application of a mid-air assistive tool, its benefit of easy-to-operate and remote interaction may allow the application of public assistive tools.

\begin{figure} [tbh!]
    \centering
    \includegraphics[scale = 0.8]{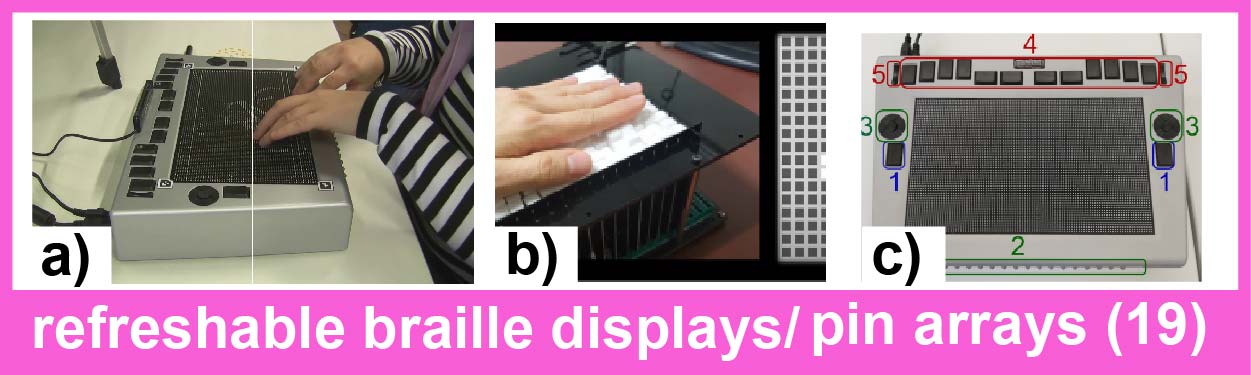}
     \caption{Examples of refreshable braille displays/pin arrays: a) A refreshable braille display that assists people with BLV's drawing \cite{bornschein2018comparing}; b) ThroughHand: a pin array system that assists people with BLV in playing 2D games \cite{Jung2021ThroughHand}; c) Audio-Tactile Reader (ATR): a multimodal refreshable braille display that assists in people with BLV's STEM education \cite{melfi2022audio}.}
    \label{fig:refreshable_braille_displays_pin_arrays}
    \Description{This figure illustrates the examples of refreshable braille displays/pin arrays, each figure includes three subfigures. Each subfigure shows one prototype.}
\end{figure}

\subsection{Underexplored Haptic Tools for Specific Tasks - Discussion for Haptic Assistive Tools for Various Tasks} \label{Haptic Assistive Tools for Understudied Application areas}
In our review of how haptic assistive devices are used for different tasks, we identified several areas with limited exploration in their applications. 
Specifically, we found no instances of haptic bands being employed to aid in understanding graphical information. 
For guidance and navigation purposes, only robot/drone haptic devices, haptic bands, white canes, and other handheld haptic devices were used. 
In the context of education and training, there was no use of haptic sliders and haptic bands. 
% Similarly, haptic gloves, robot or drone haptic devices, and white canes were not used for other life and work tasks.
These areas represent opportunities for further investigation, and we discuss potential avenues for future research in applying different haptic assistive tools for these tasks. 

\subsubsection{Underexplored haptic tools for graphical information understanding}
We found that haptic bands were not used to understand graphical information. 
% One possible explanation is that people with BLV often use their fingers to interact with haptic displays.
This may be due to the nature of the task since perceiving detailed graphical elements requires precision, which is more typically conducted through direct and tactile finger-based interactions.
These interactions allow users with BLV to perceive the variations in height across the graphs through devices like refreshable braille displays, pin arrays, and tactile graphics \cite{swaminathan2016linespace,Nagassa20233D,baker2014tactile}. 
Additionally, users with BLV rely on haptic devices to guide their fingers along the contours of the graphs. 
Examples of such guiding devices include robot or drone haptic devices and robotic-arm haptic devices \cite{guinness2019robo,ducasse2018botmap,abu2010multimodal}.
In contrast, haptic bands are typically worn on the arm, wrist, and waist, and offer guidance information with a resolution of a few centimeters. 
As a result, they may be less effective at conveying graphical information, which requires more fine-grained resolution.
However, the key function of haptic devices for graphical information understanding involves providing details about the graphs' outlines and helping people with BLV traverse and explore these graphs \cite{abu2010multimodal,fan2022slide,swaminathan2016linespace}.
Therefore, if the haptic band is small enough to be worn on the finger and convey guidance at a centimeter level, it may effectively guide the users with BLV's fingers in exploring the graphs.

In addition, only two papers in our corpus explored the application of electrotactile devices for understanding graphical information.
However, their low cost, lightweight design, flexibility, and high guidance resolution suggest potential applications in graphical information understanding, as users could explore detailed graphics for a long time \cite{jiang2024designing}.

% In addition to haptic rings, no white canes have been used to assist graphical information understanding because users commonly hold white canes to detect obstacles on the road, which requires its length to be more than one meter, even longer than the graphs to be explored \cite{nasser2020thermalcane,siu2020virtual}.
% Therefore, white canes may be unsuitable to assist users with BLV with graphical information understanding tasks.

\subsubsection{Underexplored haptic tools for guidance/navigation}
We found that only robot/drone haptic devices, haptic bands, white canes, and other handheld haptic devices were used for guidance/navigation. 
However, with further improvement, refreshable braille displays/pin arrays, tablet/smartphone integrated haptic actuators, and haptic gloves could also assist with guidance/navigation.
The potential reasons are that the haptic devices should be wearable and portable enough to be brought outdoors. 
Additionally, they should either be able to continuously guide the users with BLV relative to the current direction and location \cite{huppert2021guidecopter,hong2017evaluating} (e.g., robot/drone haptic devices, haptic bands, etc), or should be able to detect obstacles and people passing by in real-time \cite{nasser2020thermalcane,swaminathan2021from} (e.g., white canes). 
However, tactile graphics/maps are usually non-refreshable and non-zoomable, which cannot fulfill the requirement of continuous guidance and real-time obstacle detection. 
Although prior work has developed refreshable and zoomable tactile graphics using a 3D printer to print the required detailed images in real time \cite{swaminathan2016linespace}, current 3D printers are not portable and thus the users with BLV cannot wear or hold the device to guide them. 

In contrast, refreshable braille displays/pin arrays may provide continuous guidance cues by rendering the guidance information.
For example, when the users with BLV hold the device, the pins could arise and fall to form an arrow in the display that always points to the correct directions. 
In addition, it can display surrounding maps to the users with BLV that include important information, such as the direction of roads and the location of any shops and crossroads. 

In contrast to refreshable braille displays, tablet/smartphone-integrated haptic actuators may not be able to provide guidance cues because the vibrators can vibrate only along one axis. 
However, it can supplement map applications installed on tablets or smartphones. 
For example, they actuate when the users' finger moves on the road so that they will know where they should make a turn or cross the road. 

Aside from tablet/smartphone-integrated haptic actuators, haptic mice should be used with computers and thus are unsuitable for outdoor use. 
Additionally, haptic gloves have been demonstrated to provide continuous guidance cues to users with BLV by creating directional skin-stretch on the backside of the hands \cite{chase2020pantoguide}. 
Therefore, researchers can also explore how haptic gloves can assist users with BLV's guidance/navigation. 

% In comparison, haptic sliders and robotic-arm haptic devices are commonly grounded and thus cannot be used for guidance/navigation. 
3D models are non-refreshable, making them unable to provide continuous guidance cues and detect obstacles, and thus are not suitable for guidance/navigation tasks. 
In summary, with further development, refreshable braille displays/pin arrays and haptic gloves hold promise for assisting people with BLV in guidance/navigation, while tablet/smartphone integrated haptic actuators may supplement the use of map applications.

\subsubsection{Underexplored haptic tools for education/training}
Haptic sliders and haptic bands were not used for education/training. 
% The potential reason is that their form factors are rarely used to convey graphical information, which is a key part of STEM education (e.g., math, physics, biology, and chemistry) and O\&M training \cite{melfi2020understanding,petit2008refreshable}.
In this context, haptic devices should allow the teachers to edit their lecture materials easily because they typically must make repeated revisions before teaching \cite{melfi2020understanding}. 
Haptic sliders, with certain modifications, may be able to assist users with BLV in graphical information understanding, and further assist students with BLV in learning graphs and charts during class because the graphs are important in STEM education and O\&M training. 
For example, they can learn the trend of charts in math class and the map in O\&M class by moving the sliders to each location and memorizing their different sensations (e.g., different finger inclinations), which has been achieved by prior works \cite{fan2022slide,gay2021f2t}. 

% Compared to haptic sliders, haptic bands would only be capable of assisting with education/training if they can be worn on the finger and convey graphical information, which we mentioned in the second paragraph of section \ref{Haptic Assistive Tools for Understudied Application areas}. 
Compared to haptic sliders, haptic bands would be more easily applied to education/training if they could be worn on the finger and convey graphical information.
Haptic bands need to incorporate additional improvements such as reduced noise production and risk of interruption. They should also be light because the students have to use them for a prolonged period during class. 
For people with BLV, fingers are vital for perceiving the environment in daily life. Therefore, future researchers developing wearable haptic bands must carefully address how to avoid interfering with the natural functions of the fingers. 
For instance, they could adopt unobtrusive design approaches for the form factor, as demonstrated in prior work \cite{jiang2024designing, tanaka2023full}.

% In comparison, white canes have been widely used for O\&M training, though we found no papers from our literature review. The reason may be that the white canes have been used for O\&M training for a long time and are a traditional and classical method, so researchers currently focus on applying white canes to other tasks like guidance/navigation.

\section{Haptic Feedback for Various Tasks} \label{Haptic Feedback for Various Tasks}
\begin{figure} [tbh!]
    \centering
    \includegraphics[scale=0.23]{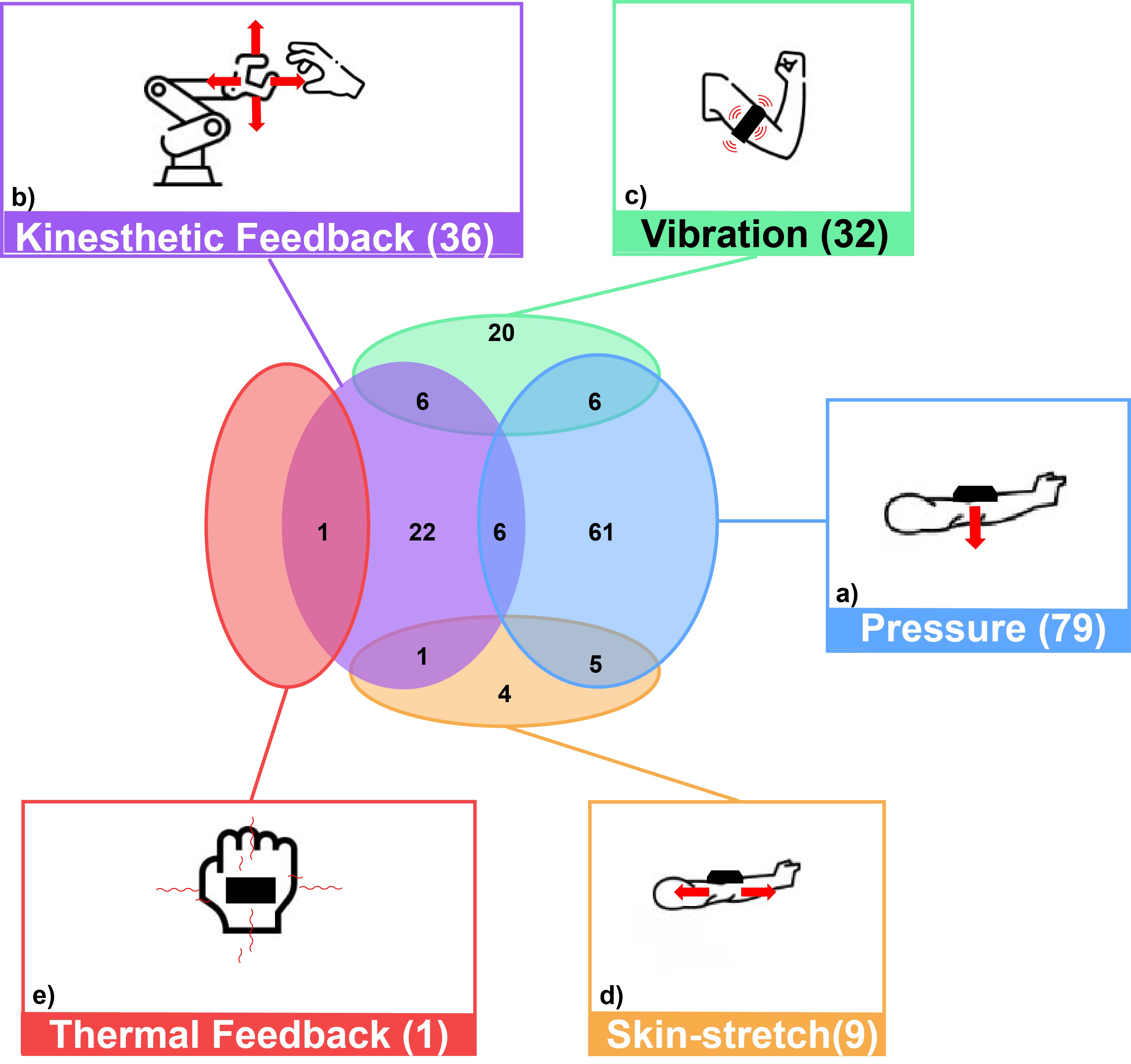}
     \caption{Examples of haptic feedback mechanisms and usage in various papers. The overlaps show a combination of various haptic feedback.}
     \Description{This figure illustrates the mechanisms of 5 haptic feedback and their relationships. The numbers represent the number of papers utilized based on the haptic feedback.}
    \label{fig:task_haptic_feedback}
\end{figure}

In addition to describing the types of haptic assistive tools that support the main tasks, we also categorized the types of haptic feedback used for various tasks. These include \textbf{pressure} (79/132), \textbf{kinesthetic feedback} (36/132), \textbf{vibration} (32/132), \textbf{skin-stretch} (9/132), and \textbf{thermal feedback} (1/132), as shown in Table \ref{table:type_of_haptic_feedback}. \textbf{Pressure} is produced by applying a normal force to the stimulation area, such as using raised lines and dots from tactile graphics/maps \cite{wang2020multimodal}, shown in Figure \ref{fig:task_haptic_feedback} a). \textbf{kinesthetic feedback} is induced by a multi-degree-of-freedom force and torque feeling related to the awareness of the position or the shape of grasped objects, such as by pulling with drones, robots, and robotic arms\cite{purves2019neurosciences}, shown in Figure \ref{fig:task_haptic_feedback} b). \textbf{Vibration} is induced by the vibration of actuators, such as voice coil actuators, linear resonant actuators, and piezo-electric actuators \cite{lim2021systematic}, shown in Figure \ref{fig:task_haptic_feedback} c). \textbf{Skin-stretch} is induced by applying tangential force to the stimulation area, such as by the motors' rotation \cite{wang2020multimodal}, shown in Figure \ref{fig:task_haptic_feedback} d). \textbf{Thermal feedback} is induced by temperature stimulation to thermoreceptors, such as by Peltiers, which is a thermal control module that has both "warming" and "cooling" effects and can quickly change the temperature \cite{nasser2020thermalcane}. They respond over a temperature range of 5 °C–45 °C, and a sharp temperature change can cause pain \cite{huang2022recent,jones2008warm}, shown in Figure \ref{fig:task_haptic_feedback} e).

\begin{table}[htbh!]
\caption{Type of haptic feedback}
\centering
\renewcommand{\arraystretch}{1.5} % Adjust row height
\small % Smaller font size
\begin{tabular}{p{2.5cm}p{3cm}p{2cm}p{3cm}p{3cm}}
\hline
\textbf{Type of Haptic Feedback} & \textbf{Graphical Information Understanding} & \textbf{Guidance /    
  Navigation} & \textbf{Education/ Training} & \textbf{Other Life and Work Tasks} \\
\hline
Kinesthetic feedback & \cite{soviak2016tactile,guinness2019robo,ducasse2016tangible,abu_doush2009making,abu2010multimodal,moll2013haptic,ducasse2018botmap,de_felice2007haptic,sallnas2007group,schloerb2010blindaid,gay2021f2t,park2015telerobotic} & \cite{siu2020virtual,huppert2021guidecopter,swaminathan2021from,guerreiro2019cabot,nasser2020thermalcane,kayukawa2020guiding,zhao2018enabling,amemiya2009haptic,amemiya2010orienting,rahman2023take,moon2019prediction,tanabe2021identification,wang2012halo,kim2015identification} & \cite{zhang2017multimodal,espinosa2021virtual,crossan2008multimodal,plimmer2011signing,murphy2015haptics} & \cite{das2023simphony,lieb2020haptic,Morelli2010vi,Plimmer2008multimodal,Park2013real} \\
\hline
Vibration & \cite{yatani2012spacesense,palani2017principles,quek2013enabling,sharmin2005first,yoo2022perception,chu2022comparative,jiang2024designing} & \cite{siu2020virtual,hong2017evaluating,erp2020tactile,Kaul2021around,kayukawa2020guiding,hirano2019synchronized,cosgun2014evaluation,lee2023novel,flores2015vibrotactile,zhao2018enabling,xu2020virtual,wang2012halo,kim2015identification,ogrinc2018sensory} & \cite{Toennies2011toward,oliveira2012the} & \cite{fink2023autonomous,sucu2014the,Stearns2016evaluating,Allman2009rock,Morelli2010vi,Cassidy2013haptic,Headley2011roughness,Rantala2009methods,Strachan2013vipong} \\
\hline
Skin-stretch & \cite{fan2022slide,kim2011handscope,gay2021f2t,chase2020pantoguide} & \cite{liu2021tactile,spiers2017design} & \cite{sanchez2010usability} & \cite{Tanaka2016haptic,kayhan2022a} \\
\hline
Thermal feedback & / & \cite{nasser2020thermalcane} & / & / \\
\hline
Pressure & \cite{shi2016tickers,swaminathan2016linespace,li2019editing,yang2020tactile,reinders2020hey,brule2021beyond,hofmann2022maptimizer,holloway20223D,fan2022slide,holloway2023tacticons,Nagassa20233D,baker2014tactile,lucentmaps2016gotzelmann,reichinger2016gesture,shi2017designing,suzuki2017fluxmarker,guinness2019robo,baker2016tactile,visually2018gotzelmann,reichinger2018pictures,prescher2017consistency,gupta2019evaluating,zeinullin2022tactile,brayda2015the,besse2018understanding,kim2015toward,braier2014haptic,mcGookin2010clutching,kim2011handscope,prescher2010tactile,kane2013touchplates,wang2009instant,ducasse2016tangible,shi2019designing,rao20202across,ducasse2018botmap,quek2013enabling,roberts2005haptic,sharmin2005first,Deschamps2012interpersonal,mascle2022tactile,levesque2012adaptive,pietrzak2009creating} & \cite{huppert2021guidecopter,barontini2021integrating,rahman2023take,spiers2017design} & \cite{albouys2018towards,holloway2018accessible,melfi2020understanding,davis2020tangible,chang2021accessiblecircuits,fusco2015tactile,holloway20193d,saikot2022refreshable,sanchez2010usability,petit2008refreshable,sargsyan20233D,oliveira2012the,brayda2013predicting,melfi2022audio} & \cite{Guo2017facade,bornschein2018comparing,Jung2021ThroughHand,das2023simphony,fink2023autonomous,bornschein2015collaborative,siu2019shapecad,baldwin2017the,Morash2018evaluating,Siu2019Advancing,Cassidy2013haptic,shi2020molder,pandey2020explore,lee2023tacnote,Headley2011roughness,Pantera2021lotusbraille} \\
\hline
\end{tabular}
\label{table:type_of_haptic_feedback}
\end{table}

\subsection{Haptic feedback for graphical information understanding}
\textbf{Pressure} is the most used type of haptic feedback in assisting graphical information understanding (43/58), followed by \textbf{kinesthetic feedback} (12/58), \textbf{vibration} (7/58), and \textbf{skin-stretch} (4/58). \textbf{Pressure} is induced by the users with BLV's touch and stroke on the raised lines and dots. The users memorize the location and shape of the lines and the combination and arrangement of a series of dots to build the images' overview gradually \cite{yang2020tactile,melfi2020understanding,shi2019designing,soviak2016tactile,brayda2015the,fan2022slide,guinness2019robo,yatani2012spacesense}. 

\begin{figure} [tbh!]
    \centering
    \includegraphics[scale = 0.8]{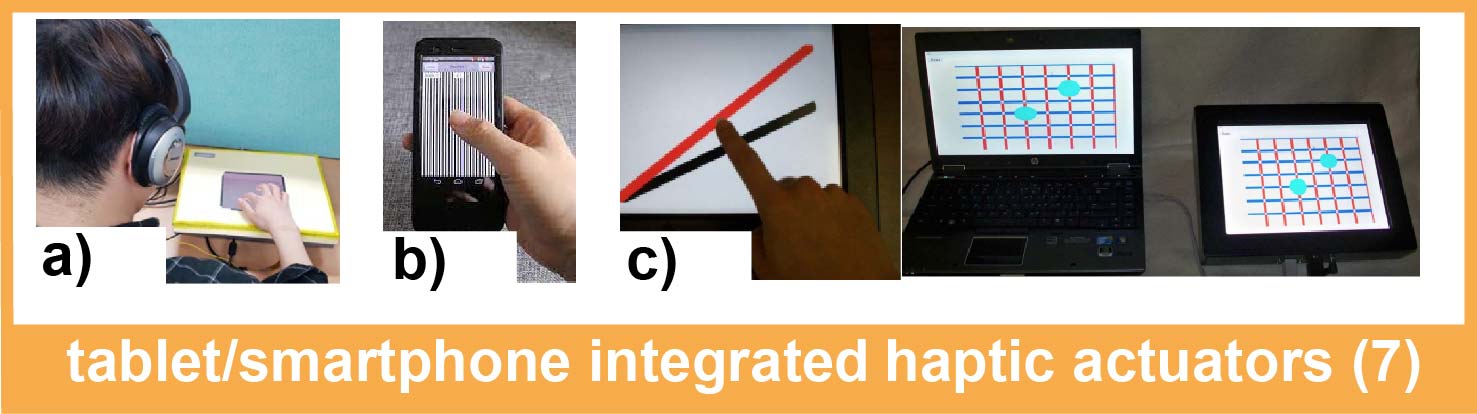}
     \caption{Examples of tablet smartphone integrated haptic actuators: a) tablet/smartphone integrated haptic actuator system that assists BLpeople with BLVnteracting with touchscreens wased on vibration \cite{chu2022comparative}; b) A tablet/smartphone haptic device that conveys graphical and mathematical concepts through vibration and audio cues \cite{Toennies2011toward}.}    
     \Description{This figure illustrates the examples of tablet smartphone integrated haptic actuators, each figure includes three subfigures. Each subfigure shows one prototype.}
     \label{fig:tablet_smartphone_integrated_haptic_actuators}
\end{figure}

\textbf{Vibration} rendered graphical information by vibrating when the user's fingers touched the lines of graphs presented on the smartphone or tablet's screen. The users with BLV build an overview of the graphs by continuously feeling and memorizing the locations and connections of the lines. These vibrators could either be integrated with the mobile device \cite{Toennies2011toward} shown in Figure \ref{fig:tablet_smartphone_integrated_haptic_actuators} c), or attached to an environmental device \cite{yatani2012spacesense,sharmin2005first}.

\subsection{Haptic feedback for guidance/navigation}
\textbf{Kinesthetic feedback} (14/26) and \textbf{vibration} (14/26) are the most used haptic feedback in assisting guidance/navigation, followed by \textbf{pressure} (4/26), \textbf{skin-stretch} (2/26), and \textbf{thermal feedback} (1/26). Robots and drones leveraged \textbf{kinesthetic feedback}  \cite{huppert2021guidecopter} or integrated with other feedback to guide users with BLV when they walk along a predefined trajectory  \cite{nasser2020thermalcane}. 

\textbf{Pressure} was integrated with \textbf{skin-stretch} and \textbf{kinesthetic feedback} to provide haptic direction cues to guide users when walking \cite{barontini2021integrating,spiers2017design}. \textbf{Thermal feedback} assists people with BLV's guidance by changing the temperature and application direction to indicate go or stop and help them find their way  \cite{nasser2020thermalcane}.

\subsection{Haptic feedback for education/training}
\textbf{Pressure} (14/20) and \textbf{kinesthetic feedback} (5/20) are the most used haptic feedback in supporting education/training, followed by \textbf{vibration} (2/20) and \textbf{skin-stretch} (1/20). \textbf{Pressure} mainly assists the education of graphical-related knowledge, such as geometry\cite{petit2008refreshable,melfi2022audio} and geography \cite{albouys2018towards,holloway20193d,holloway2018accessible}. One study used the combination of pressure and skin-stretch to improve students with BLV' navigation skills \cite{sanchez2010usability}. Robotic arms leverage \textbf{kinesthetic feedback}  to guide users with BLV's fingers to provide an intuitive understanding of charts and 2D shapes in math education\cite{abu2010multimodal,espinosa2021virtual}, biology \cite{murphy2015haptics}, and children's handwriting \cite{plimmer2011signing}. 

\textbf{Vibration.} When users with BLV touched lines and shapes on the touchscreens, a vibration was applied to help them recognize shapes \cite{Toennies2011toward,palani2017principles}. In addition, \textbf{vibration} was integrated with pressure to assist students with BLV by focusing on the whiteboard the instructor is pointing at, thus bringing them back to mainstream classrooms \cite{oliveira2012the}. 

\subsection{Haptic feedback for other life and work tasks}
\textbf{Pressure} (16/28) and \textbf{vibration} (9/28) are the most used haptic feedback in assisting other life and work tasks, followed by \textbf{kinesthetic feedback} (3/28) and \textbf{skin-stretch} (2/28). \textbf{Pressure} is applied to assist users with BLV with hand drawing/writing \cite{bornschein2018comparing,pandey2020explore}, interacting with appliances \cite{Guo2017facade,baldwin2017the}, gaming \cite{Jung2021ThroughHand}, weaving \cite{das2023simphony}, driving \cite{fink2023autonomous}, assistive tool design \cite{bornschein2015collaborative,shi2020molder}, 3D modeling \cite{siu2019shapecad}, braille reading \cite{Morash2018evaluating,Pantera2021lotusbraille}, annotation \cite{lee2023tacnote}, and roughness perception \cite{Headley2011roughness}. 

\textbf{Vibration} was leveraged by vibrators to provide direction cues in assisting drivers with BLV \cite{sucu2014the} and improve the game experience \cite{Strachan2013vipong,Morelli2010vi,Allman2009rock}. \textbf{kinesthetic feedback} was provided by robotic arms' guidance to assist people with BLV with 3D modeling \cite{lieb2020haptic}, hand drawing/writing, and telepresence \cite{Park2013real}. \textbf{Skin-stretch} is induced by people with BLV moving the haptic slider for audio production \cite{Tanaka2016haptic}.

\subsection{Exploring Haptic Feedback: Usage, Comparisons, and Future Applications - Discussion for Haptic Feedback for Various Tasks}
Our findings demonstrated how five types of haptic feedback were used for different tasks, and we further discuss each haptic feedback in this section. 
We first describe the utilization of various haptic feedback in different tasks, including graphical information understanding, guidance/notification, education/training, and other tasks (e.g., cooking and baking). We then make comparisons with other haptic feedback to explain their respective strengths and limitations. 
Finally, we discuss the potential applications of each haptic feedback for future research.

\subsubsection{Potential usage of kinesthetic feedback}
Kinesthetic feedback is induced when mechanoreceptors within the joints and muscles detect the body's movement or halt due to force applied by actuators. 
Our literature review showed that kinesthetic feedback has been widely used for all tasks. 
people with BLV utilize kinesthetic feedback to perceive the movement of their arms, aiding in understanding the contours of 2D graphs, shapes of 3D models, and directional cues provided by robotic-arm haptic devices and robot/drone haptic devices \cite{lieb2020haptic,zhang2017multimodal,espinosa2021virtual,huppert2021guidecopter,guinness2019robo}. 
Alternatively, they sense abrupt halts in hand movement caused by force applied to the hands, aiding in obstacle detection while using white canes \cite{nasser2020thermalcane,xu2020virtual,swaminathan2021from}.
Compared to pressure and skin-stretch, kinesthetic feedback necessitates precise force application to either facilitate or impede body movement \cite{park2020effect}. 
Therefore, while kinesthetic feedback can offer clear haptic cues due to its effective force, it requires more powerful actuators that are typically larger and heavier \cite{see2022touch}.
% Despite these drawbacks, in principle, kinesthetic feedback can be employed in any task benefiting from haptic guidance cues and notifications. 
Despite these drawbacks, kinesthetic feedback may be employed in tasks that involve haptic guidance cues and notifications.
For instance, in education/training, it can assist users with BLV in painting or sculpting by guiding their hands along painting or cutting trajectories. 
In baking, it can facilitate ingredient search and addition step by step. 

\subsubsection{Potential usage of vibration}
Vibration occurs when mechanoreceptors under the skin detect the vibrating actuators. 
Our literature review indicates widespread use of vibration across tasks, except for education/training. 
people with BLV perceive guidance cues generated by haptic bands' vibrations for navigation or gaming \cite{hong2017evaluating,erp2020tactile,flores2015vibrotactile}, or they sense notifications while touching lines and graphs on smartphones \cite{yoo2022perception,chu2022comparative}. 
Unlike other haptic feedback methods, vibration is cost-effective, requiring lower power consumption, and its low-profile vibrators can be discreetly attached under clothing or surfaces.
However, it may not offer as precise haptic cues as kinesthetic feedback \cite{ooka2010virtual}. 
Vibrators attached in different locations produce varying direction cues, with closer placements providing higher resolution. 
Yet, kinesthetic feedback, utilizing motor rotation, offers substantially higher resolution, around 0.18 degrees, surpassing vibration \cite{kaneko1989spherical}.
Nevertheless, despite its limitation in rendering high-resolution haptic guidance cues, vibration still finds utility in education/training. 
For instance, it can simulate object weight to aid children with BLV in understanding weight differences. 
This simulation is achieved through vibrotactile phantom sensation (VPS), precisely controlling vibration magnitude and vibrator location. 
Maintaining constant magnitude while adjusting vibrator distance generates tangential force while maintaining magnitude and moving vibrators back and forth generates slip \cite{ooka2010virtual}.  

\subsubsection{Potential usage of skin-stretch}
Skin-stretch arises when sheer force activates mechanoreceptors. 
Our literature review uncovered fewer than five instances where skin-stretch was employed across all four tasks. 
Users with BLV either perceived skin-stretch through the movement of haptic sliders to grasp graphic contours \cite{fan2022slide,kim2011handscope}, or they sensed it from motors' rotation, providing continuous guidance cues \cite{liu2021tactile,spiers2017design}. 
Both tasks rely on haptic guidance cues provided by skin-stretch, although its effectiveness is somewhat limited without the ability to vary haptic intensities, unlike other types of feedback.
However, skin-stretch offers guidance with compact, lightweight mechanical structures, typically employing just a servo motor, making it portable. 
We found that skin-stretch devices are suitable for on-the-go guidance/navigation, supported by an example from Liu et al. \cite{liu2021tactile}.
Skin-stretch devices excel in indoor O\&M training due to their compactness and lightness.
When combined with vibration, skin-stretch devices can offer both haptic guidance and notifications mirroring real BLV usage scenarios.

% Unlike traditional training tools like white canes, which only detect barriers, skin-stretch devices, when combined with vibration, offer haptic guidance and notifications mirroring real BLV usage scenarios (e.g., using white canes alongside smartphone guidance).

\subsubsection{Potential usage of thermal feedback}
Thermal feedback occurs when temperature changes activate thermal receptors. 
Our investigation uncovered only one instance of thermal feedback used in guidance/navigation, with no instances in the other tasks.
people with BLV perceive guidance cues through thermal feedback stimulation from peltiers. 
Unlike other haptic feedback types, thermal feedback is susceptible to environmental temperature and clothing effects \cite{halvey2011effect,halvey2012baby}.  
However, it offers versatility in conveying both haptic guidance and notification cues across various applications.
In cooking and baking, it may provide food temperature information, offering a more intuitive experience and protecting individuals with BLV from burns compared to audio notifications, which was discussed in Section \ref{Haptic Assistive Tools for Understudied Application areas}.
For graphical information understanding, wearing peltiers on fingertips can signify different map areas through temperature changes. 
For example, when they touch buildings, the temperature rises to represent the indoor environment, and when they touch the street or other outdoor locations, the temperature falls to represent the outdoor environment. 
For education/training, thermal feedback may aid with understanding objects or creatures. 
For instance, a temperature drop could indicate cold-blooded animals like snakes, while a rise could suggest warm-blooded ones like rabbits.

\subsubsection{Potential usage of pressure}
Pressure feedback is generated when normal force stimulates mechanoreceptors.
We found fewer than five instances of pressure used in guidance/navigation.
Users with BLV commonly perceive height variations across graphs through tactile graphics/maps, refreshable braille displays/pin arrays, haptic mice, and gloves \cite{swaminathan2016linespace,Jung2021ThroughHand,Nagassa20233D,siu2019shapecad,prescher2017consistency}, or sense shapes of 3D models, robot/drone haptic devices, handheld devices, and haptic sliders \cite{shi2016tickers,huppert2021guidecopter,fan2022slide,brayda2015the,goncu2010usability}. 
% Pressure is the most widely used haptic feedback among all types.
% Unlike other haptic devices, pressure devices can be as simple as tactile maps or 3D models, making them lightweight and easy to create, hence their widespread use. 
% However, pressure is less utilized than kinesthetic feedback and vibration for guidance/navigation. 
% This is partly due to larger, heavier pressure guidance devices compared to vibration-based ones and lower guidance accuracy (0.18 degrees, achieved by motor rotation) compared to kinesthetic feedback. 

% Therefore, pressure is better suited for other tasks. 

Pressure is one of the most commonly used forms of haptic feedback. 
Unlike many other haptic technologies, pressure-based devices can range from simple tactile maps to 3D models, which makes them relatively lightweight and easy to produce, contributing to their broader application. 
However, pressure was used less frequently than kinesthetic feedback and vibration for guidance and navigation purposes. 
This is partly because pressure-based guidance devices tend to be larger and heavier compared to vibration-based alternatives, and they offer lower guidance accuracy (0.18 degrees, achieved through motor rotation) in comparison to kinesthetic feedback. 
As a result, pressure may be applied to other types of tasks that simulate the perception of holding objects.
For instance, haptic gloves can simulate holding virtual chess or laser guns, enabling people with BLV to play chess and First-Person Shooter (FPS) games without visual cues \cite{qi2023haptglove}.   
By mimicking the sensation of holding chess pieces or laser guns, players with BLV can engage in virtual gaming with spatial awareness cues provided by sound direction and intensity changes.

\section{On-body Stimulation Position for Tasks} \label{Stimulation Position for Tasks}
\begin{figure} [tbh!]
    \centering
    \includegraphics[scale=0.65]{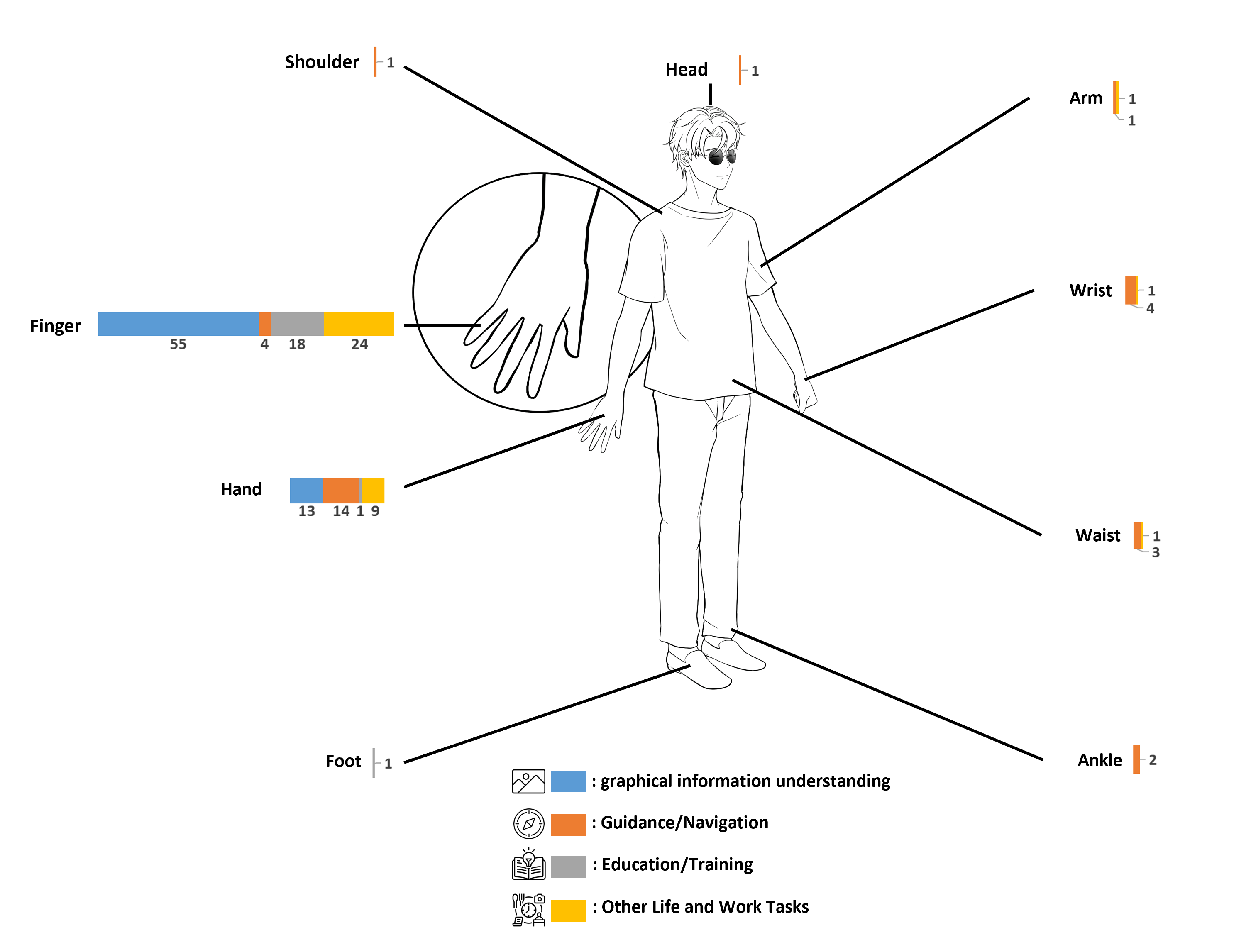}
     \caption{BLV tasks achieved by stimulating different on-body stimulation positions: the bars illustrate the frequency of each on-body stimulation position utilized for different tasks.}
     \Description{This figure illustrates the on-body stimulation positions pointed to a human's body figure. The on-body stimulation positions include finger, hand, wrist, waist, ankle, arm, shoulder, foot, and head. The bars near each on-body stimulation position illustrate the frequency of each on-body stimulation position utilized for different tasks.} 
     \label{fig:stimulation_position}
\end{figure}

Haptic assistive tools produced haptic feedback on eight on-body stimulation positions, which include \textbf{finger} (101/132), \textbf{hand} (37/132), \textbf{wrist} (5/132), \textbf{waist} (4/132), \textbf{ankle} (2/132), \textbf{arm} (2/132), \textbf{shoulder} (1/132), \textbf{foot} (1/132), and \textbf{head} (1/132). Figure \ref{fig:stimulation_position} shows the distribution of people with BLV's tasks in each on-body stimulation position.

\subsection{Finger} 
The \textbf{finger} is most used for graphical information understanding (55/101), followed by other life and work tasks (24/101) and education/training (18/101), and is least used for guidance/navigation (4/101). 

Users with BLV used their \textbf{finger} to touch and stroke the raised lines, dots, and shapes of the tactile graphics/maps, refreshable braille displays/pin arrays, and 3D models. The pressure induced by touching and stroking helped them build an overview of the graphical information and continuously update this overview \cite{shi2016tickers,petit2008refreshable}. Users with BLV could also sense vibrations when touching images on the smartphone's touchscreen and gradually explored the border and shape of the images \cite{palani2017principles,yoo2022perception}. 
Other haptic feedback induced on the \textbf{finger} include skin-stretch from a slider and mouse \cite{fan2022slide,kim2011handscope,gay2021f2t} and kinesthetic feedback from robotic arms to guide their \textbf{finger} along the predefined trajectory of images, which assisted users with BLV in understanding their content \cite{park2015telerobotic,schloerb2010blindaid}.

For other life and work tasks, the \textbf{finger} perceived the relative height changes of refreshable braille displays/pin arrays' actuators or pins to build the 3D models and to read braille characters \cite{Guo2017facade,Siu2019Advancing,Morash2018evaluating}. 
The \textbf{finger} could also perceive the vibration from actuators attached to a haptic band or haptic mouse to generate roughness sensations in graphs or to indicate the impact of a ball in a haptic game \cite{Stearns2016evaluating,Headley2011roughness,Strachan2013vipong}. 
Furthermore, the kinesthetic feedback from the robotic arms guided the \textbf{finger} along a predefined trajectory \cite{Plimmer2008multimodal,Park2013real}. 

To assist with education/training, \textbf{finger} perceived pressure from various 3D models to distinguish the map's features for O\&M training and the structure of a circuit board based on differences in shapes \cite{chang2021accessiblecircuits,fusco2015tactile}, and kinesthetic feedback from robotic arms to help students explore the 3D shapes in class and learn handwriting \cite{espinosa2021virtual,plimmer2011signing}. 

\begin{figure} [tbh!]
    \centering
    \includegraphics[scale = 0.8]{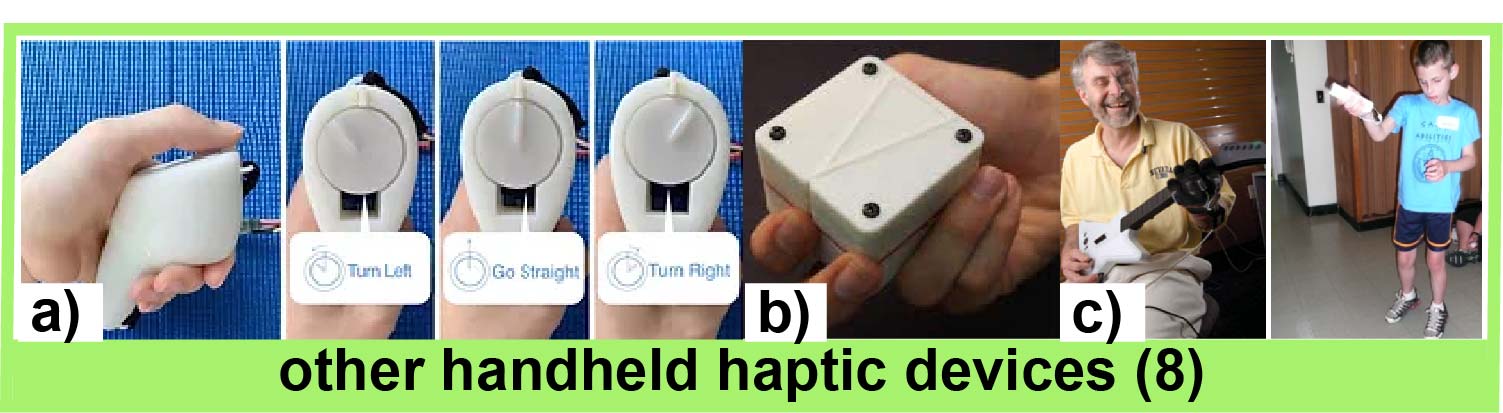}
     \caption{Examples of other handheld haptic devices: a) Tactile Compass: a holdable haptic device that guides people with BLV to travel independently based on the skin-stretch guidance cues \cite{liu2021tactile}; b) A shape-changing haptic device that conveys direction cues for people with BLV's navigation \cite{spiers2017design}; c) VI-Bowling: a holdable haptic device for people with BLV's gaming. The figure shows how users with BLV leverage the device to play Blind Hero and VI Tennis (two games) \cite{Morelli2010vi}.}
     \Description{This figure illustrates the examples of other handheld haptic devices, each figure includes three subfigures. Each subfigure shows one prototype.}
    \label{fig:other_handheld_haptic_devices}
\end{figure}

\begin{figure} [tbh!]
    \centering
    \includegraphics[scale = 0.8]{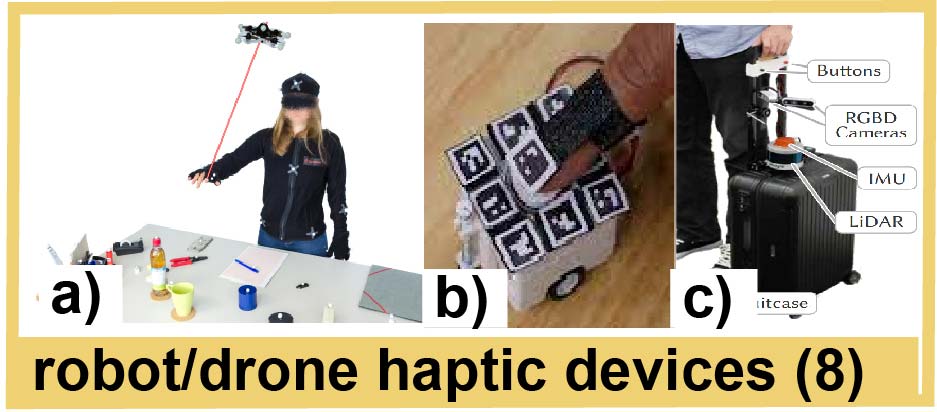}
     \caption{Examples of robot/drone haptic devices: a) GuideCopter: a drone-based haptic guidance interface to assist users with BLV in finding objects \cite{huppert2021guidecopter}; b) A guiding system based on tablet robots to assist people with BLV in finding objects on the table \cite{rahman2023take}; c) A guiding system that assists people with BLV to walk in public spaces and to avoid pedestrians \cite{kayukawa2020guiding}.}
     \Description{This figure illustrates the examples of robot/drone haptic devices, each figure includes three subfigures. Each subfigure shows one prototype.}
    \label{fig:robot_drone_haptic_devices}
\end{figure}

To assist guidance/navigation, \textbf{finger} perceived pressure and skin-stretch from the pull of drones and robots \cite{huppert2021guidecopter,rahman2023take} shown in Figure \ref{fig:robot_drone_haptic_devices} a) and b), and the deformation of other handheld haptic devices \cite{liu2021tactile, spiers2017design}, shown in Figure \ref{fig:other_handheld_haptic_devices} a) and Figure \ref{fig:other_handheld_haptic_devices} b).

\subsection{Hand} 
The \textbf{hand} is the most used for guidance/navigation (14/37), graphical information understanding (13/37), other life and work tasks (9/37), and education/training (1/37). Compared to the \textbf{finger} that can sense finer haptic cues such as pressure and skin-stretch by touching and stroking in a small area, the \textbf{hand} has a larger contact area with the haptic assistive tools on its palmer and backside. 

Haptic assistive tools mainly applied kinesthetic feedback on the \textbf{hand} to assist with guidance/navigation, where the \textbf{hand} perceived kinesthetic feedback by holding a white cane to sense the obstacles on the road \cite{nasser2020thermalcane,zhao2018enabling} and by following the guidance of the robots and drones via holding their handles, to walk along the streets \cite{rahman2023take,guerreiro2019cabot}. 

To assist with graphical information understanding, the hands' palmer side was used as a supplement to fingers to strengthen the sensation of lines and shapes \cite{reinders2020hey,brule2021beyond}. The \textbf{hand} also sensed the vibration from a series of vibrators attached to the back of the smartphone that were distributed in a matrix shape. Each vibrator represented one place on the map and the users could understand the spatial relationships between these places by sensing the difference of vibrating locations \cite{yatani2012spacesense}.

To assist with other life and work tasks, haptic assistive tools also applied pressure and vibration on the whole \textbf{hand}, such as the pressure from a mid-air haptic device that presented the map of a street intersection by applying different pressure on the palm \cite{fink2023autonomous}
Another example includes pressure from a pin array system in a tactile game. The game required the players to sense moles and hit them with finger presses. The sensation of moles was created by pressure because of the pins' rise \cite{Jung2021ThroughHand}.

\subsection{Other on-body stimulation positions} The other on-body stimulation positions were used in fewer than or equal to five tasks. 
The \textbf{wrist} was used to assist with guidance/navigation (4/5) and other life and work tasks (1/5), specifically during gaming via vibration from the vibrators attached to the haptic bands \cite{Allman2009rock,grussenmeyer2017evaluating,lee2023novel,siu2020virtual,ogrinc2018sensory}. The \textbf{waist} was used to assist guidance/navigation (3/5) and other life and work tasks (1/5) with haptic bands attached to the \textbf{waist}, like in the form of a belt.
The vibrators were uniformly attached to the haptic bands in an equal adjacent distance. 
The direction cues were provided by vibrating the vibrators in the appropriate direction \cite{erp2020tactile,cosgun2014evaluation,flores2015vibrotactile,kayhan2022a}. 
Specifically, one work applied this technique to assist runners with BLV during their running process \cite{kayhan2022a}. 
The \textbf{arm} was used to assist guidance/navigation (1/2) and other life and work tasks (1/2), specifically during gaming via pressure, skin-stretch, and vibration induced by vibrators and motors from haptic bands \cite{barontini2021integrating,Allman2009rock}.
The haptic bands worn on the \textbf{arm} were all on the upper \textbf{arm}, which was a different location than the haptic bands worn on the wrist. 
In particular, one paper leveraged both \textbf{arm} and \textbf{wrist} to convey vibration cues to indicate targets to be hit in a game. 
The vibration of the \textbf{arm} indicated the farther target, while the vibration of the wrist indicated the nearer target \cite{Allman2009rock}. 
The \textbf{ankle} (2/2) and the \textbf{shoulder} (1/1) were used to assist guidance/navigation via vibration induced by haptic bands\cite{hirano2019synchronized,xu2020virtual}.
Xu et al. asked the participants with BLV to wear a backpack that was attached to four vibrators in three arrangements. 
All the arrangements included the attachment of two fixed vibrators on the \textbf{waist}, while the other two vibrators were attached to the \textbf{shoulder}, the \textbf{ankle}, and the \textbf{wrist} individually. 
The fixed two vibrators indicated left and right direction while the other two vibrators on the \textbf{waist} indicated go and stop. 
They compared the guiding performance of these three arrangements and found that attaching the other two vibrators on the \textbf{shoulder} could present the highest guidance accuracy \cite{xu2020virtual}. 
The \textbf{foot} (1/1) was used to assist with education. 
Sanchez et al. developed a grounded haptic assistive tool that assist blind children in learning navigation skills by sensing the pressure and skin-stretch from the device that indicated the direction cues \cite{sanchez2010usability}. 
The \textbf{head} (1/1) assisted people with BLV's navigation by using 24 vibrators around the \textbf{head}. 
The device could notify the users to move and stop and provide direction cues by vibrating different groups of vibrators. 
They found that, their device could achieve an average absolute deviation of 5.7 cm  from the predefined trajectories \cite{Kaul2021around}. 

\subsection{Rarely-used On-body Stimulation Positions for Various Tasks - Discussion for On-body Stimulation Position for Tasks} \label{Rarely-used Stimulation Positions For Various Tasks}
The results of how on-body stimulation positions are used for different tasks showed that certain on-body stimulation positions are rarely used in applications.
Specifically, our findings indicate that the finger and hand are the most commonly employed on-body stimulation positions, whereas the wrist, waist, ankle, arm, shoulder, foot, and head are utilized in only five instances or less.
In the subsequent discussion, we explore the potential factors contributing to this pattern and suggest possible avenues for future research.

\subsubsection{Two-point Discrimination of Various On-body Stimulation Positions}
Previous research has utilized a metric known as successive two-point discrimination to ascertain the distribution of mechanoreceptors across different regions of the human body. 
This metric measures the minimal distance at which two stimuli can be distinguished when applied successively \cite{dellon1987reliability}. 
A lower two-point discrimination value indicates a higher density of mechanoreceptors and increased sensitivity to position.
The finger exhibits the highest distribution of mechanoreceptors (0.3 cm), followed by the hand (0.4 cm for the palm, 0.9 cm for the dorsum), head (0.6 cm for the forehead), foot (0.6 cm for the sole, 1.4 cm for the dorsum), waist (1.2 cm for the back), arm (1.5 cm for the forearm), wrist (1.7 cm), and shoulder (1.9 cm)  \cite{koo2016two,mancini2014whole}.
Our literature review reveals that the utilization of various on-body stimulation positions in the examined papers partially aligns with the recognized distribution of mechanoreceptors across the human body. 
Specifically, the fingers and hands emerge as the most sensitive on-body stimulation positions and are consequently prevalent in various tasks.
However, despite the head and foot being more sensitive than the wrist, waist, arm, and shoulder, they are underutilized on-body stimulation positions. 
One explanation could be the ease of wearing haptic assistive tools on the wrist, waist, arm, and shoulder compared to the foot. 
Additionally, wearing such tools on the head may attract unwanted attention and may be less favored by users with BLV.

\subsubsection{Potential Applications of Various On-body Stimulation Positions}
Although these areas are less frequently utilized, they present potential for further exploration. 
For instance, researchers could design haptic assistive tools to stimulate different areas under the sole to provide feedback mimicking various terrains such as deserts, snowfields, or rocky lands as mentioned by Wang et al. \cite{wang2018design}. 
Apart from the head and foot, other on-body stimulation positions such as the arm, wrist, waist, shoulder, and ankle are primarily utilized for wearing haptic bands for guidance/navigation but can be leveraged for additional tasks. 
For instance, haptic assistive tools could enhance the film-watching experience for individuals with BLV by providing pressure and thermal feedback worn on these positions to mimic whole-body sensations, such as being bitten by a snake or feeling wind and fire in front of the users, as demonstrated by Delazio et al. through pneumatic devices \cite{delazio2018force}. 
This haptic feedback enriches the film-watching experience for individuals with BLV by compensating for information lost due to lack of vision.  
Compared to traditional methods like audio descriptions, haptic assistive tools seamlessly integrate into the film-watching process without disrupting dialogue, sound effects, or musical scores \cite{viswanathan2011enhancing}. 
In summary, various on-body stimulation positions offer potential for diverse applications, although they are currently underexplored.

\section{Limitations of Haptic Assistive Tools and User Studies} \label{Limitations of Haptic Assistive Tools and User Studies}
Although haptic assistive tools have been designed to assist people with BLV in their daily lives and work, they still contain many limitations. The main categories include \textit{hardware limitations}, \textit{functionality limitations}, and \textit{UX and evaluation method limitations}.
Under each category, we further divided the limitations into subcategories. 
% Figure \ref{fig:hardware_limitations}, Figure \ref{fig:functionality_limitations}, and Figure \ref{fig:user_limitations} summarizes the main categories, subcategories, and the associated limitations under each subcategory. 

\subsection{Hardware limitations} 
\label{Hardware limitations}
Hardware limitations include \textbf{environmental effects on sensors and actuators}, \textbf{conspicuous appearance and noise}, and \textbf{cumbersome and heavy devices}.
% which are shown in Figure \ref{fig:hardware_limitations}. 
\textbf{Environmental effects on sensors and actuators} describe how environmental effects, such as physical issues and user habits, affected the sensors' sampling quality and haptic perception. \textbf{Conspicuous appearance and noise} describe how the haptic assistive tools' conspicuous appearance provoked stress in users with BLV and how the noise negatively affected the users with BLV's perception of haptic cues. \textbf{Cumbersome and heavy devices} describe how haptic assistive tools' large size and heavy weight affected their daily use and caused inconvenience for users with BLV. 

% \begin{figure} [tbh!]
%     \centering
%     \includegraphics[scale=0.35]{figure/Hardware_limitation.png}
%      \caption{Hardware limitations include three categories: external effects on sensors and actuators, conspicuous appearance and noise, and cumbersome and heavy devices.}
%      \Description{This figure illustrates the hardware limitations, including external effects on sensors and actuators, conspicuous appearance and noise, and cumbersome and heavy devices.}
%     \label{fig:hardware_limitations}
% \end{figure}

\textbf{Environmental effects on sensors and actuators} include \textit{inaccurate sensors and haptic perception due to physical issues} and \textit{imprecise haptic perception due to users' habits}. Physical issues are induced by the change in physical conditions related to the haptic assistive tools, such as the deformation of users' skin and the orientation of wearable devices around the arm because of long time wear. Some haptic devices leveraged optical sensors, such as RGB-D cameras, to recognize on-the-road obstacles and avoid them. However, these cameras vibrated when the users moved, which lowered the sensors' sampling quality by introducing signal noise. This issue impacted the application of haptic cues, common for haptic bands, robot/drone haptic devices, and other handheld haptic devices \cite{kayhan2022a}. In addition, physical issues also affected users with BLV's haptic perception. 
For example, while a robot guided users, it used vibrations to warn them of obstacles. However, vibrations from the robot were hard to distinguish from those caused by the rough surface of the floor, leading to confusion \cite{kayukawa2020guiding}.

Additionally, the quality of sensor sampling and users' haptic perception were influenced by their habits. 
For example, when users wore haptic bands on the wrist for guidance while using a white cane, how they held the cane changed the relative angles of their hand to the ground. 
If the coordinates of the guiding system were based on the haptic bands rather than the ground, the direction cues would be based on the haptic bands' coordinates.
This discrepancy meant that the cues did not match the actual directions needed in the real-world, leading to inaccurate guidance \cite{kayhan2022a}. 

As noted in prior work, sensor inaccuracy may lead to safety issues, such as collisions in crowded places or falls due to inaccurate guidance on surfaces with height differences \cite{thiyagarajan2022intelligent}. 
Given the critical nature of safety, additional precision in navigation is necessary.
% As mentioned by prior work, sensors' inaccuracy may cause safety issues, such as collision in a crowded place and falling due to inaccurate guidance to the surface with height differences \cite{thiyagarajan2022intelligent}.
% As the safety issue is critic, additional precision for navigation is necessary.

\textbf{Conspicuous appearance and noise} includes \textit{noise from actuators} and \textit{conspicuous haptic assistive tools}. Motors and vibrators produced noise during their operation, affecting users' perception of haptic cues and audio reports of multimodal assistive tools \cite{huppert2021guidecopter,Kaul2021around}. Specifically, Kaul et al. developed a head-worn haptic assistive tool for people with BLV's guidance. However, the noise from the vibrators negatively affected people with BLV's recognition because some vibrators were close to their ears. They reported that, after bone conduction, the noise was especially loud and possibly disruptive, distracting them from the direction cues \cite{Kaul2021around}. 

The conspicuous appearance of haptic assistive devices, such as robots/drones and other tools exceeding the size of a smartphone, negatively impacted users with BLV. 
The use of these conspicuous tools drew unwelcome attention and discussion among nearby individuals and made users with BLV feel uncomfortable, as they preferred not to be perceived differently due to their visual conditions \cite{kayhan2022a,guerreiro2019cabot}. 
In comparison, users with BLV tend to feel less self-conscious if assistive devices resemble mainstream products \cite{shinohara2011in}.
Moreover, users with BLV also have aesthetic preferences for their assistive tools---a consideration that is often overlooked by current manufacturers \cite{shinohara2011in}.

\textbf{Cumbersome devices} include \textit{Inconvenience of grounded and heavy devices}. Due to their large size and weight, stemming from their mechanical designs, robotic-arm haptic assistive tools and haptic sliders were typically mounted on a table or secured to the ground.
These tools limited user mobility, for example, teachers had to transport these heavy devices across classrooms, or students with BLV had to relocate to the specific classroom equipped with the device. This arrangement was inconvenient for both teachers and students with BLV \cite{murphy2015haptics,crossan2008multimodal}.

Furthermore, the size and weight of robots and drones made them impractical for everyday use, and they could pose safety risks, particularly in crowded or narrow spaces. For instance, Kayukawa et al. found that users with BLV felt this equipment was ``too large and heavy for daily use,'' expressing concerns about the potential for collisions with other pedestrians in densely populated areas \cite{kayukawa2020guiding}.

\subsection{Functionality limitations}
\label{Functionality limitations}
Functionality limitations include \textbf{lack of information display control}, \textbf{lack of detailed information}, and \textbf{few presentation modalities}.
% which are shown in Figure \ref{fig:functionality_limitations}. 
The \textbf{lack of information display control} describes the drawbacks of limited control of their haptic assistive tools and limited refreshing and zooming functions on people with BLV's image sensemaking process. The \textbf{lack of detailed information} describes the difficulties in presenting complex structure and geometry, and how the limited display resolution of graphical information fails to present complex images to the users with BLV. \textbf{Few presentation modalities} describe how only using a single type of haptic feedback may affect haptic assistive tools' information convey performance.

% \begin{figure} [tbh!]
%     \centering
%     \includegraphics[scale=0.35]{figure/Functionality_limitation.png}
%      \caption{Functionality limitations include three categories: lack of information display control, lack of detailed information, and few presentation modalities.}
%      \Description{This figure illustrates the functionality limitations, including lack of information display control, lack of detailed information, and few presentation modalities.}
%     \label{fig:functionality_limitations}
% \end{figure}

\textbf{Lack of information display control} includes \textit{lack of user control of assistive tools} and \textit{non-refreshable and non-zoomable images}. The lack of user control of assistive tools includes the control of speed and intensity of haptic feedback. On one hand, the speed for haptic sliders and robotic-arm haptic assistive tools is usually fixed. This is to control the number of variables for the user study. However, users wish to adjust the speed as they may be unable to follow a fast-speed display and find a low-speed display tedious. On the other hand, the lack of control of haptic feedback intensity may induce unclear display and potential safety issues. People have various thresholds and endurance of specific haptic feedback \cite{jiang2021douleur,lin2022super}. Low intensity may create haptic cues that the user cannot perceive, while a high intensity may induce undesired stress and potential harm to the users, as the faster the haptic assistive tools' movement, the larger the force applied to the users with BLV. 
If participants were unaccustomed to the movement speed, there was a risk that these devices could unexpectedly contact their fingers or hands. Such incidents could not only startle the users but also potentially cause injury \cite{abu2010multimodal,fan2022slide}. 

Tactile graphics and maps, typically produced in print form, cannot update or zoom in on images, making them less time and cost-effective compared to dynamic haptic assistive technologies like refreshable braille displays. 
Moreover, without the capability to zoom, people with BLV cannot access detailed levels of information within graphics, preventing them from focusing on specific areas of interest. Access to these details is crucial for the sense-making process of users with BLV, thus the absence of such detailed insight can hinder their ability to fully grasp the content of the images \cite{swaminathan2016linespace}.

\textbf{Lack of detailed information} includes \textit{difficulty in presenting complex structures and geometry} and \textit{limited display resolution of graphical information}. Current tactile graphics/maps typically present a single layer of image, chart, or map. However, it is hard for users with BLV to comprehend their complex indoor structures with several separated layers of tactile maps for buildings with more than one floor. In this case, a multistorey building structure should also be presented to assist users with BLV with building an overview. Lack of such presentation may induce inefficient indoor navigation within a complex building, such as in an airport. In addition, some tactile maps did not provide different haptic cues to display various types of paths and areas, which resulted in users with BLV's hardness in differentiating private areas from public areas and pedestrian streets from roads for vehicles. Lack of such differentiation may induce safety issues such as trespassing. How to make the above-mentioned differentiation with simple presentations is important because it can reduce users' short-term memory load of memorizing the maps and reduce their risks of undesired trespassing \cite{hofmann2022maptimizer}.

Some refreshable braille display/pin arrays suffered from low presentation resolutions, which decreased the presented images' details. Low presentation resolutions can only assist users with BLV in building an overview of simple shapes or images. However, current STEM education requires the image presentation of complex structures, such as biology and geography, which is not achievable by some of the refreshable braille display/pin arrays. The lack of such presentation may induce misunderstanding of biological structures, such as the cells and landscapes important for biology and geography \cite{melfi2022audio}.

\textbf{Few presentation modalities} include \textit{lack of information due to single type of haptic feedback}. It is hard to convey multidimensional information for assistive tools that only provide one type of haptic feedback. For example, Espinosa et al. leveraged a robotic-arm haptic device to assist students with BLV in learning the 3D models with the guidance of a robotic arm. However, the device can only present the models' 3D outlines and cannot render the models' texture only with kinesthetic feedback. This reduced the realism and immersion when people with BLV touched the 3D models and thus failed to assist them in learning the textural features of various objects. \cite{espinosa2021virtual}.

\subsection{UX and evaluation method limitations}
\label{UX and evaluation method limitations}
UX and evaluation method limitations include \textbf{lack of user group diversity consideration}, \textbf{lack of customization}, \textbf{limitations on user study}, and \textbf{haptic device distrust}.
% , which are shown in Figure \ref{fig:user_limitations}. 
The \textbf{lack of user group diversity consideration} describes how the lack of consideration of users with various visual conditions, users who have different assistive tools usage habits, and users of various ages and education backgrounds affected the design and use of haptic assistive tools. The \textbf{lack of customization} describes how people with BLV's user experience was affected because they could not customize their 3D models. \textbf{Limitations on user study} describes how the lack of comparison with other haptic devices, variability in the use of subjective rating scales, and lack of real-world experiments affected the design of haptic assistive tools and user studies. \textbf{Haptic device distrust} described that the users with BLV distrusted the robots and drones because of their unfamiliarity with these haptic assistive tools.

% \begin{figure} [tbh!]
%     \centering
%     \includegraphics[scale=0.35]{figure/User_limitation.png}
%      \caption{Functionality limitations include four categories: lack of user group diversity consideration, lack of customization, limitations on user study, and haptic device distrust.}
%      \Description{This figure illustrates the UX and evaluation method limitations, including lack of user group diversity consideration, lack of customization, limitations on user study, and haptic device distrust.}
%     \label{fig:user_limitations}
% \end{figure}

\textbf{Lack of user group diversity consideration} includes \textit{lack of visual condition considerations}, \textit{lack of assistive tools usage habits consideration}, and \textit{lack of user ages and education background consideration}. Many haptic assistive tools were designed specifically for totally blind users, though there are more low-vision people in the world than totally blind people. Prior works have discussed their limitations: to make more users with BLV use their haptic assistive tools, adding features such as sharp contrast colors for low-vision users is necessary \cite{lee2023tacnote,holloway2023tacticons}. In comparison, a lack of consideration of the people with BLV's different visual conditions may exclude potential users from using the haptic assistive tools. 

Some handheld haptic devices interfered with existing assistive tools as they required users to hold the device with their dominant hand, which was typically used to hold white canes and manage guide dogs. 
Some current haptic assistive tools only have a guidance function, and users must hold the white canes or guide dogs with their non-dominant hand to recognize the obstacles \cite{liu2021tactile,spiers2017design}.
Adopting haptic assistive technologies required these users to modify their routine activities, invest time, and adapt to new techniques. The learning curve associated with these tools, along with the increased risk of bumping into obstacles during the adjustment phase, should be addressed. 

Users with various ages and educational backgrounds affected the design of haptic devices. Elders and children cannot use or wear a device for a long time due to their health conditions. Therefore, the weight should be considered when designing a handheld or wearable haptic assistive tool \cite{kayukawa2020guiding,holloway2023tacticons}. In addition, the tools designed for elders, children, and users with low education backgrounds should be easy to use as their learning efficiency is commonly lower than other users \cite{liu2021tactile,Morelli2010vi}.  Though not every work aims to provide a one-size-fits-all solution, broadening the user groups from adults only to minorities such as elders, children, and people with low education backgrounds allows us to gain a deeper understanding of their practical needs for the improvement of future prototypes.

\textbf{Lack of customization} includes \textit{Users' dissatisfaction with predefined 3D models}. The users were dissatisfied with the 3D models that did not consider the cultural and religious differences of users. For example, the Red Cross, Red Crescent, and Red Crystal were used in different regions around the world, which were all the emblems of the International Red Cross and Red Crescent Movement (IRRC) \cite{redcross2021}. However, designers without such knowledge overlooked these discrepancies and were unfamiliar with more regional, cultural, or religious symbols. Lack of such consideration may induce misunderstanding and even bias among the users \cite{holloway20223D}. In addition, users with BLV require time to learn how to recognize predefined 3D icons. For instance, individuals from different generations are familiar with objects relevant to their own times but may not recognize items associated with other eras \cite{holloway2023tacticons}. The absence of customized 3D models can substantially strain short-term memory, forcing users to memorize new 3D models, particularly of objects they are not familiar with, such as specific map landmarks \cite{holloway20223D,holloway2023tacticons}.

\textbf{Limitations on user study} include \textit{lack of comparisons with other haptic devices}, \textit{variability in use of subjective rating scales}, and \textit{lack of real-world experiments}. A lack of comparisons with other haptic devices may prevent researchers from understanding the effectiveness of their devices with a known baseline. 
In contrast, studies that include comparisons can reveal specific advantages. 
For example, Fan et al. found that their interactive haptic assistive tools were more intuitive than tactile graphics and saved time compared to traditional 3D models, thanks to their refreshable nature \cite{fan2022slide}. 
Without such comparisons, it becomes difficult to determine if other haptic assistive tools are competitive in performance, making it harder for future researchers to build on their work.
% Fan et al. compared their interactive haptic assistive tools with tactile graphics and 3D models to understand their line charts display performance. They found that their devices provided position and inclination cues that could assist users with BLV in building an overview more intuitively than tactile graphics. Compared to 3D models, their devices provided a refreshable line chart display, which saves time for producing new 3D models \cite{fan2022slide}. 
% While for the other works that did not compare their haptic assistive tools with the other tools, it is hard for us to ensure that their devices are competitive in performance, and thus are hard to follow by future researchers.

The variability in the use of rating scales among the studies presents a challenge in evaluating the overall user experience of these tools. 
For instance, van Erp et al. assessed mental effort using the NASA-TLX scale, which focuses on cognitive workload \cite{erp2020tactile}, while Albouys et al. employed the System Usability Scale (SUS) and the User Experience Questionnaire (UEQ) to gauge user satisfaction, without directly measuring workload \cite{albouys2018towards}. 
The inconsistency in evaluation metrics complicates comparisons across studies, making it difficult to draw meaningful conclusions about the comprehensive user experience.

Conducting research in real-world settings enhances the ecological validity of a study, offering insights into the practical use of haptic assistive tools by users with BLV \cite{baxter2015chapter}. The absence of real-world experiments might fail to uncover the specific challenges, which results in overlooked design and usability problems. This impacts the overall effectiveness of the tools developed for users with BLV \cite{hofmann2022maptimizer,melfi2020understanding}.

\textbf{Haptic device distrust} is shown by users' \textit{distrust of robots and drones due to unfamiliarity}. 
Users with BLV often experience distrust towards robots and drones due to their unfamiliarity with such technology, which differs substantially from the tools they use daily. For those who have never interacted with robots or drones, understanding their appearance and functionality is challenging, leading to feelings of insecurity. In particular, participants are concerned about collisions with robots or injuries from drone propellers, affecting both the users and bystanders \cite{kayukawa2020guiding,rahman2023take}. Furthermore, Huppert et al. reported that initial worries about drone safety were prevalent among participants, who required time to adjust to using them \cite{huppert2021guidecopter}. The challenge of ensuring drone safety was identified as a major limitation of their system, underscoring the need for additional research in this area.
\section{Discussion}
We conducted a review of papers from the last twenty years to evaluate the use of haptic assistive tools, haptic feedback mechanisms, and on-body stimulation positions in supporting individuals with BLV across various tasks. Our findings, detailed in Section \ref{Haptic Assistive Tools for Various Tasks}, reveal predominant reliance on tactile graphics/maps in graphical information understanding (23/132) and on haptic bands in guidance/navigation (10/132). For education/training, tactile graphics/maps and robotic-arm haptic devices each accounted for (5/132) of the studies, while other life and work tasks frequently utilized refreshable braille displays/pin arrays (8/132).

Section \ref{Haptic Feedback for Various Tasks} highlights that pressure feedback is extensively used in graphical information understanding (43/132), with kinesthetic feedback and vibration each supporting guidance/navigation and education/training (14/132). Pressure was the main type of feedback for other life and work tasks (16/132).
In Section \ref{Stimulation Position for Tasks}, our analysis shows that the fingers (101/132) and hands (37/132) are the most common on-body stimulation positions.
Section \ref{Limitations of Haptic Assistive Tools and User Studies} identified three primary limitations within current haptic assistive tools: hardware constraints, functionality issues, and UX and evaluation method limitations.

Moving forward, our discussion will address the implications of our findings from several key perspectives. 
Initially, we will identify areas where haptic assistive tools can be improved.
Additionally, we will explore opportunities for developing new assistive devices based on current haptic technologies.
Subsequently, we will examine the relationship of haptic and audio feedback. 
Finally, we will discuss the limitations of our systematic literature review.

% we will assess both the challenges effectively mitigated by existing haptic assistive tools and those that remain unresolved. 

\subsection{Areas of Improvement for Existing Haptics Assistive Tools \label{Areas of Improvement}}
Section \ref{Limitations of Haptic Assistive Tools and User Studies} has revealed various limitations of existing haptic assistive tools. 
We provide improvement suggestions in three main categories to address these limitations: hardware improvements, functionality improvements, and UX and evaluation method improvements. 
Under each category, we further divided the improvements into subcategories. 
Figure \ref{fig:hardware_improvement}, Figure \ref{fig:functionality_improvement}, and Figure \ref{fig:user_improvement} summarize the main categories, subcategories, and the associated improvements under each subcategory.

\subsubsection{Hardware improvements} 
\begin{figure}[tbh!]
    \centering
    \includegraphics[scale=0.5]{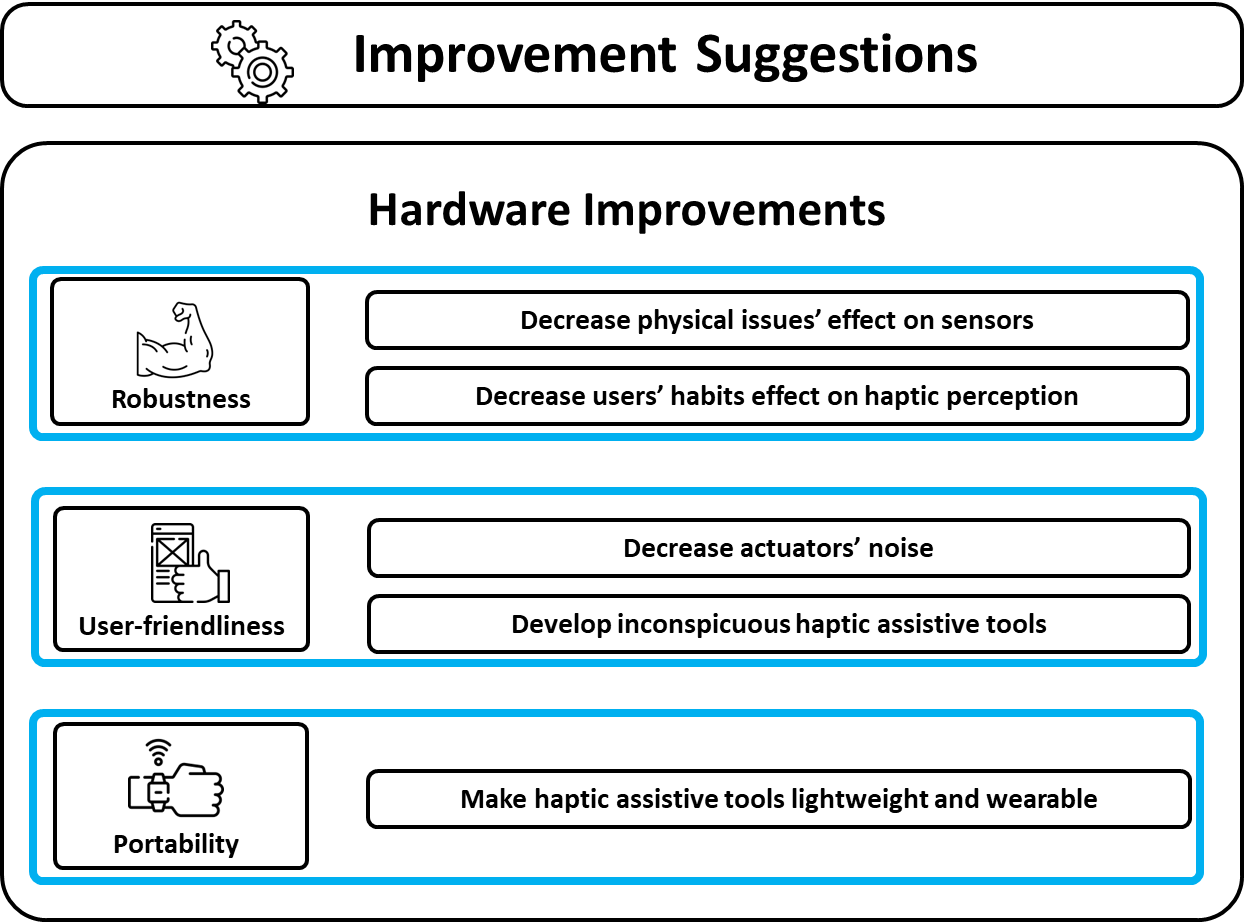}
     \caption{Hardware Improvement Suggestions include three categories: robustness, user-friendliness, and portability}
      \Description{This figure illustrates the hardware improvements, including robustness, user-friendliness, and portability}
    \label{fig:hardware_improvement}
\end{figure}

The hardware of haptic assistive tools requires enhancements in the following areas:  \textbf{robustness}, \textbf{user-friendliness}, \textbf{portability}. 
Detailed suggestions for achieving these attributes are outlined in the following paragraphs, which are shown in Figure \ref{fig:hardware_improvement}.

\textbf{Robustness} is achieved by \textit{decreasing physical issues' effect on sensors} and \textit{decreasing users' habits effect on haptic perception}.  

\textit{Decreasing physical issues' effect on sensors}:  
To decrease the effect of physical issues such as the deformation of users' skin, and the orientation of wearable devices around the arm because of long-term wear, on the sensors, both hardware and software design aspects must be addressed. 
In hardware design, ensuring a secure attachment of the haptic devices to the user's body is crucial. 
Some devices lack proper attachment mechanisms, leading to unintended movement and rotation during use, which negatively affects sensor sampling quality \cite{kayhan2022a,liu2021tactile,kayhan2022a}. 
Utilizing leather belts instead of cloth ones can offer greater stability due to their higher friction, preventing shifting or rotation over prolonged wear periods. 
On the software side, signal filtering techniques are essential for processing sensor data accurately. 
The choice of filtering method should be informed by the specific external environment and types of noise present. 
For instance, a median filter effectively eliminates salt-and-pepper noise by replacing the current value with the signal windows' median values.
Conversely, a Gaussian filter is effective for smoothing continuous signals by applying weighted averaging based on a Gaussian function \cite{orfanidis1995introduction,zhang2022modern}. 
By employing appropriate filtration algorithms, the sensor signals required for activating haptic feedback can be refined accurately, ensuring optimal performance of haptic assistive tools.

\textit{Decreasing users' habits effect on haptic perception}: 
Considering users' habits is essential in the design of haptic assistive tools. 
For instance, when users of haptic bands alter their grip on the white cane, it can affect the relative angle between the haptic device and the ground. 
To address this, researchers can integrate IMU sensors or optical trackers to detect such changes in real-time. 
By aligning the guiding system's coordinates with the haptic bands, this approach ensures that guidance cues are accurately provided, mitigating the risk of misleading directions.

\textbf{User-friendliness} is enhanced through strategies such as \textit{decreasing actuators' noise} and \textit{developing inconspicuous haptic assistive tools}. 

\textit{Decreasing actuators' noise}:
Different types of actuators generate varying levels of noise. 
Researchers could address this by selecting actuators that inherently produce less noise or by integrating external noise-dampening solutions. 
For instance, adding soundproof or vibration-insulating layers, such as lightweight soundproof foam, could effectively reduce the noise caused by actuators \cite{sui2015lightweight}.

% The more powerful the actuator is, the stronger the haptic feedback perceived by the users, but that comes at the cost of higher noise. 
% Researchers can incorporate external soundproof or vibration insulation layers, such as soundproof foam, to mitigate the noise produced by actuators \cite{sui2015lightweight}. 

\textit{Developing inconspicuous haptic assistive tools}: 
To ensure inconspicuousness, devices should deliver haptic cues discreetly, without drawing unnecessary attention \cite{boldini2021inconspicuous}.
This aligns with the "Design for Social Accessibility" approach, which emphasizes the importance of both social and functional factors in assistive technology design \cite{shinohara_incorporating_2018, shinohara_design_2020}.
For instance, devices can be concealed behind users' clothing or seamlessly integrated with everyday items like watches or smartphones \cite{nasser2020thermalcane,hong2017evaluating}.

\textbf{Portability} is attained through \textit{making haptic assistive tools lightweight and wearable}.
Research findings highlight the inconvenience of cumbersome devices in various life and work scenarios. 
To overcome this challenge, haptic assistive tools should prioritize portability by utilizing wearable designs or integrating diverse haptic actuators into existing handheld haptic tools like white canes or smartphones. 
For example, vibrators and peltiers have proven effective in providing directional cues, making them suitable for seamless integration into white canes  \cite{drebushchak2008peltier,nasser2020thermalcane, siu2020virtual}. 

Vibrators attached to smartphones can facilitate graphical information understanding, such as interpreting images and maps \cite{palani2017principles}. Moreover, alternative haptic feedback types can be explored to deliver desired cues. 
While kinesthetic feedback is commonly used to guide finger exploration of graphs, traditional robotic-arm haptic devices providing such feedback are bulky and stationary \cite{abu2010multimodal,lieb2020haptic}. 
In contrast, a recent study employing electrotactile feedback to guide finger exploration highlighted its advantage of being lightweight \cite{jiang2024designing}. 
Our review identified only two full papers that specifically utilized electrotactile feedback, while other related work on using electrotactile guidance for touchscreen interactions was limited to extended abstracts \cite{xu2011tactile, kajimoto2014hamsatouch}. 
This gap presents an opportunity to further explore the design of assistive technologies leveraging electrotactile feedback to enhance their portability and performance.

Besides graphical information understanding, researchers can leverage skin-stretch to provide guidance instead, as demonstrated by Chase et al. \cite{chase2020pantoguide}. They offered a lighter device that, with both skin-stretch-based and kinesthetic feedback devices, can effectively support users with BLV's comprehension of charts.

\subsubsection{Functionality improvements} 
\begin{figure}[tbh!]
    \centering
    \includegraphics[scale=0.5]{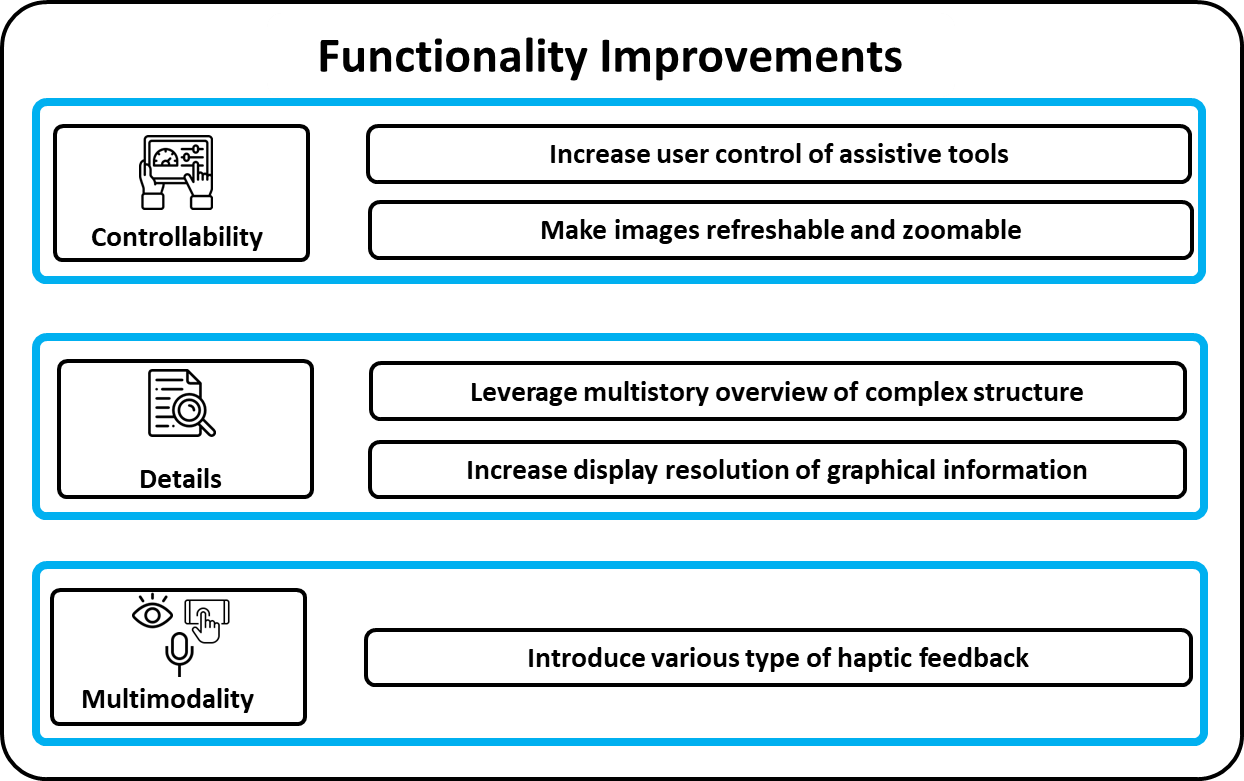}
     \caption{Functionality Improvement Suggestions include three categories: controllability, details, and multimodality}
     \Description{This figure illustrates the functionality improvements, including controllability, details, and multimodality}
    \label{fig:functionality_improvement}
\end{figure}

The functionality of haptics assistive tools can be improved in the following areas: \textbf{controllability}, \textbf{details}, and \textbf{multimodality}. We provide suggestions in the following paragraphs on how to achieve these attributes, which are shown in Figure \ref{fig:functionality_improvement}

\textbf{Controllability} is achieved by \textit{increasing user control of assistive tools}, and \textit{making images refreshable and zoomable}. 

\textit{Increasing user control of assistive tools}: 
Allowing users control over devices like robotic-arm haptic devices, haptic sliders, robot/drone haptic devices, and haptic bands can improve their willingness to use these tools and enhance information acquisition efficiency. 
Researchers can implement two methods to enable user control of their devices.

First, designers can incorporate buttons that users can traverse using their screen readers.
These buttons can be labeled "low/small," "medium," and "high/large" to control display speed or haptic feedback intensity, or they can use numeric values (ranging from 0 to 5, where 0 signifies stop and 5 indicates maximum speed or intensity).  
This design facilitates quick initiation and simplifies the learning process for users. 

Second, researchers can integrate voice-based interactions, allowing users to adjust display speed and haptic feedback intensity via voice commands.
Voice commands may offer finer control, particularly suitable for users who desire precise adjustments \cite{potluri2021examining,lang2021pressing}. 
Recent advancements in large language models enable users to issue vague commands like "make this quicker" or "lower the intensity", with the models autonomously predicting and executing the desired adjustments \cite{kuzdeuov2024chatgpt,wu2023brief}. 
These interactions are more natural, allowing even novice users to gain swift control of their haptic assistive tools. 

\textit{Making images refreshable and zoomable}: 
Introducing interactive components such as tablet robots and 3D models can make tactile graphics/maps' information refreshable. 
Tablet robots, a type of mobile robot small enough to navigate on a table, can be paired with tactile graphics to provide dynamic updates \cite{guinness2019robo}. 
These robots, along with 3D models serving as icons or landmarks, can modify the information presented by the tactile graphics through movement and changes in their positions. 
For instance, tablet robots moving across a map can simulate changes in vehicle volume across different city areas, while alterations in icons can signify shifts in species distribution within a small area or even a specific continent \cite{guinness2019robo,shi2016tickers}.

Furthermore, enhancing tactile graphics/maps' usability involves making them zoomable, a feature achievable through integration with 3D printers.
By combining a 3D printer with a camera capable of recognizing blank areas on a whiteboard, detailed portions of graphics/maps can be printed based on users' voice input \cite{swaminathan2016linespace}. 
However, it is important to note that this approach may incur higher \rv{time and financial} costs. Users must wait for the printing and cooling process, which can take anywhere from a few minutes to over an hour, and they are also responsible for the ongoing consumption of 3D printing materials.
% However, it is essential to note that this approach may incur higher costs timely and financially: the users have to wait during the printing and cooling process (from a few minutes to tens of minutes) and have to pay for the continuous consumption of 3D printing materials. 
Thus, researchers should carefully consider the cost implications when adopting or developing this technique.

\textbf{Details} can be enhanced by \textit{leveraging multi-storey overview of complex structure} and \textit{increasing display resolution of graphical information}. 

\textit{Leveraging multi-storey overview of complex structure}: 
A multi-storey overview can be achieved by integrating various layers of tactile graphics/maps into a single system with a sliding or rotating design, as demonstrated by Nagassa et al. \cite{Nagassa20233D}. 
In sliding designs, each layer of tactile maps is individually fixed to a series of parallel sliders.
These layers overlap, and each can be moved sideways with its corresponding slider. 
This overlapping arrangement offers an integrated overview of how each floor is interconnected, including the locations of elevators and escalators. 
Users can then move each layer sideways to explore the detailed structure of each floor.

Rotation designs involve fixing each layer of tactile maps to a rotation pod, with layers overlapping similarly to sliding designs. 
The overlap provides an integrated overview, and each layer can be rotated to different angles to reveal detailed single-floor structure information \cite{Nagassa20233D}. 
Compared to using separate tactile maps for each floor, these designs allow users to acquire detailed information about each floor while simultaneously building an integrated overview of the overall structure. 
This approach helps strengthen users' understanding of complex structures while reducing the cognitive load associated with processing vast amounts of information.

\textit{Increasing display resolution of graphical information}:  
A new mechanical design is necessary to improve the resolution of refreshable braille displays/pin arrays. 
Current resolutions range from 4 to 6 mm, substantially higher than the 0.5 mm resolution of human SA-I mechanoreceptors, which detect sustained pressure \cite{sekerere2022investigation,kandel2000principles,johnson2001roles}. 
To achieve higher resolution, new actuators like microneedles and electrotactile actuators can be integrated. 
Microneedles offer point-to-point resolutions as low as 30 $\mu$m, surpassing the resolution of SA-I mechanoreceptors \cite{miranda2023hollow} (the closer the distance between adjacent actuators, the higher the resolution).
Additionally, electrotactile displays can achieve a resolution of 1.15 mm, which is suitable for conveying graphical information at a resolution lower than SA-I mechanoreceptors \cite{lin2022super}. 
This enhancement in resolution ensures more precise tactile feedback, enabling users to discern finer details in tactile graphics and text.

\textbf{Multimodality} is achieved by \textit{introducing various types of haptic feedback}, enriching the information conveyed to users with BLV. 
Robotic-arm haptic devices, for instance, can offer vibration and kinesthetic feedback to relay various cues, including location and guidance information. 
In a chart comprehension system developed by Abu-Doush et al., vibration intensity denoted the current traverse area, while kinesthetic feedback directed the user's finger along a predetermined chart trajectory \cite{abu2010multimodal}. 
Additionally, thermal feedback can complement tactile graphics/maps by distinguishing indoor and outdoor areas. 
For instance, higher temperatures signify indoors, while lower temperatures indicate outdoors.
By integrating different haptic feedback modalities, such as vibration, kinesthetic, and thermal cues, graphical information like maps can be conveyed more effectively and intuitively.

\subsubsection{UX and evaluation method improvements} 
\begin{figure}[tbh!]
    \centering
    \includegraphics[scale=0.5]{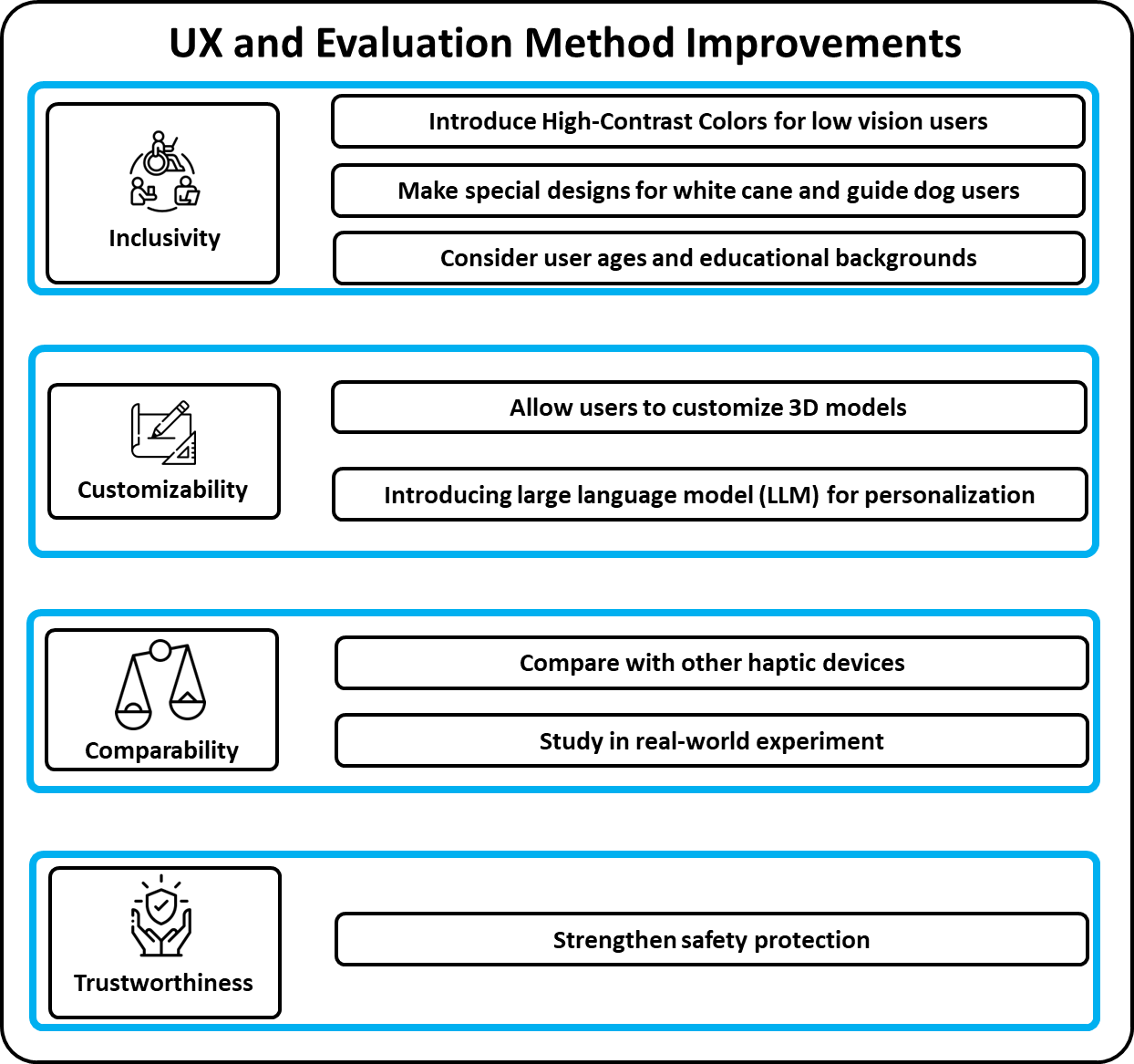}
     \caption{User Improvement Suggestions include four categories: inclusivity, customizability, compatibility, and trustworthiness}
     \Description{This figure illustrates the UX and evaluation method improvements, including inclusivity, customizability, compatibility, and trustworthiness}
    \label{fig:user_improvement}
\end{figure}

The UX and evaluation methods of haptics assistive tools could benefit from enhancements in several key areas: \textbf{inclusivity}, \textbf{customizability}, \textbf{comparibility}, and \textbf{trustworthiness}. We provide suggestions in the following paragraphs on how to achieve these attributes, which are shown in Figure \ref{fig:user_improvement}.

\textbf{Inclusivity} is achieved by \textit{introducing high-contrast colors for low vision users}, \textit{making special designs for white cane and guide dog users}, and \textit{considering user ages and educational backgrounds}. 

\textit{Introducing high-contrast colors for low vision users}: 
High-contrast colors are particularly beneficial for low-vision people, as they still retain residual vision. 
Color contrast refers to the difference in brightness between different colors, such as black text on a white background or vice versa.
Previous research has highlighted the effectiveness of combining haptic and visual cues to assist low-vision people in comprehending graphical information. 
For example, incorporating high-contrast colors into tactile graphics and visual elements can help users distinguish different parts of figures and maps.
Utilizing a color contrast checker during the design process can further aid users in discerning graphical details \cite{zimmerman2011blindness,holloway2023tacticons}. 
By integrating haptic feedback with high-contrast colors, haptic assistive tools can enhance the efficiency of information acquisition for low-vision users.

\textit{Making special designs for white cane and guide dog users}: 
When designing new haptic assistive tools, it is important to consider the integration of existing tools like white canes and guide dogs. 
Traditionally, users holding a white cane or accompanied by a guide dog have their dominant hand occupied, posing a challenge when using handheld haptic assistive tools that also require the dominant hand. 
This forces users to adapt their habits to accommodate both tools.
To address this issue, researchers could explore the development of wearable haptic assistive tools as an alternative to handheld devices.
Unlike handheld devices, wearable haptic tools, such as haptic bands worn on the wrist, waist, shoulder, arm, or foot, offer promising solutions. 
Previous studies have demonstrated the effectiveness of wearable haptic devices in conveying guidance cues, expanding their potential beyond traditional handheld devices \cite{sucu2014the,lee2023novel}. 

\textit{Considering user ages and educational backgrounds}: 
When designing haptic assistive tools, it is crucial to consider users' ages and educational backgrounds to ensure usability.
Lightweight design is paramount to accommodate users across various demographics.
Components like light DC motors, as demonstrated by Liu et al. in their tactile compass development \cite{liu2021tactile}, and flexible peltiers used in devices like the ThermalCane by Nasser et al. \cite{nasser2020thermalcane}, along with flexible printed circuits as seen in the super-resolution electrotactile haptic device by Lin et al. \cite{lin2022super}, contribute to the overall lightness of the devices.
In addition, drones, which do not require physical contact, are another option for lightweight haptic assistive tools, though safety measures need to be enhanced for elderly or child users \cite{rahman2023take,abu2010multimodal}. 
In addition to being lightweight, haptic assistive tools should feature intuitive interfaces and clear instructions suitable for users with varying levels of technological proficiency. 
For instance, advanced functions that require fine-grained control can be hidden initially, with basic functions readily accessible.
This layered approach to control helps reduce confusion during the initial learning phase, particularly for elderly, children, or less technologically savvy users \cite{darejeh2013review,haikio2007touch}.

\textbf{Customizability} is achieved by \textit{allowing users to customize 3D models} and \textit{introducing AI-powered personalization}

\textit{Allowing users to customize 3D models. }
Different cultures and eras may have unique symbols for representing features on maps, such as religious symbols or icons denoting gender-neutral bathrooms.
Some countries used bi-gender symbols, while others utilized a symbol that merges the features of various genders to indicate its all-gender characteristic. 
Ignoring these differences can lead to misunderstandings or conflicts. 
Co-designing with individuals with BLV can help ensure that 3D models align with users' cultural and contextual backgrounds, as demonstrated in previous research \cite{mussi2021co}. 
Co-design processes can leverage 3D printers and large language models (LLMs) to facilitate collaboration. 
For instance, LLMs can automatically generate 3D models based on user descriptions, as illustrated by Jun et al. and Liu et al.\cite{jun2023shape,liu2024one}.
Users can then iterate on these models, gradually becoming more familiar with their designs and enhancing their ability to distinguish between different features \cite{kuzdeuov2023chatgpt,sun20233d}. 

\textit{Introducing large language models (LLM) for personalization.} 
The various features of haptic feedback were typically controlled by the researchers, which limited users with BLV's independent use of haptic assistive tools as researchers and volunteers were not always available \cite{jiang2024designing,fan2022slide}.
Traditional control methods required the users to understand complicated control steps and conduct slow operations to avoid discomfort from the haptic feedback, thus was slow and reduced user experience \cite{abu2010multimodal}.

Recent advances in LLMs, such as ChatGPT, have shown
their potential to assist users with BLV in various areas, including understanding graphical information \cite{zhao2024vialm,hwang2024safe,yang2024viassist}, education \cite{bahar2023chatgpt,enkelejda2023chatgpt}, and navigation \cite{hwang2024safe,wang2024visiongpt}.
Inspired by its ability to understand the users' various semantics, LLMs might be used to understand the users' requirements to help them adjust the haptic feedback's features, such as force intensity, vibration frequency, and temperature level.
Moreover, customized haptic patterns might also be created for users with BLV, such as notifications of various barriers for haptic bands, and representations of the figures' different areas for refreshable braille displays/pin arrays.

\textbf{Comparability} is achieved by \textit{comparing with other haptic devices} and \textit{studying in real-world experiments}. 

\textit{Comparing with other haptic devices}: 
When comparing haptic assistive tools with other works, researchers should focus on key parameters relevant to the tasks they aim to assist users with BLV in completing more efficiently. 
These parameters may vary depending on the task at hand.
For instance, when evaluating the performance of haptic assistive tools for graphical information understanding, researchers may measure participants with BLV's accuracy in constructing mental maps by tasks such as drawing or describing figures or charts. 
However, measuring mental models through tactile drawing can be challenging due to people with BLV's limited exposure to drawing and the lack of holistic feedback \cite{kamel2000study}.
Thus, these measurements should be used with caution.
Specific features of the charts can also be compared to assess the extent to which users with BLV comprehend them \cite{fan2022slide,abu2010multimodal}. 
Similarly, in evaluating the efficacy of haptic assistive tools for navigation/guidance, researchers may measure standardized variables such as differences in angles and deviations from predefined routes \cite{liu2021tactile}. 
By comparing these key parameters, researchers can gain objective insights into the advantages and limitations of their devices compared to prior works.

\textit{Studying in real-world experiments}: 
Researchers should consider how the environment may impact haptic assistive tools in real-world experiments. 
For instance, temperature variations between indoor and outdoor environments can impact thermal feedback. 
In colder settings, the relative temperature difference between on-body stimulation positions and surrounding skin areas increases, potentially enhancing the perception of thermal stimuli \cite{nasser2020thermalcane,boulant1986temperature,taylor2011human}.
This temperature sensitivity should be considered when using thermal feedback intensity as a notification mechanism.
In addition, some participants wear thicker clothes and coats while being outdoors. 
For haptic devices worn on the wrist, waist, and arm, participants' clothing may affect the thermal feedback, skin-stretch, and pressure applied by the haptic assistive tools. 
Conversely, laboratory studies afford control over indoor temperatures and clothing conditions, allowing researchers to minimize such environmental influences on haptic feedback. 
While conducting research in real-world environments demands more effort and consideration, it presents invaluable opportunities to identify the practical limitations of haptic assistive tools and refine them for optimal real-world use.

\textbf{Trustworthiness} is achieved by \textit{strengthening safety protections}. 
Users should have a clear awareness of the device's operational status and the ability to override its functions promptly and easily. 
For instance, users should be empowered to halt haptic cues instantly if they prefer not to be guided, whether through a voice command, a simple click, or by removing their fingers from the actuators. 
Wearable devices, where actuation cessation might not be as straightforward, should feature easily releasable belts, enabling users to withdraw their hands swiftly \cite{rahman2023take}. 

Furthermore, ensuring the safety of drones or robots is paramount to fostering user trust. 
Protective cages can mitigate potential harm from rotating propellers, as demonstrated by Rahman et al. and Abtahi et al. \cite{rahman2023take,abtahi2019beyond}. 
Additionally, Guerreiro et al. ensured that the robots moved ahead and kept a constant distance from the users with BLV to avoid possible collision \cite{guerreiro2019cabot}. 
Implementing these protective measures not only enhances user safety but also cultivates trust in the reliability and usability of haptic assistive tools. 
When users feel in control of the devices' operation and assured of their safety, their confidence and willingness to engage with the tools are heightened.

In addition, we found similarities between the UX improvements that we synthesized with those suggested by prior work \cite{schneider2017haptic,kim2020defining}
Schneider et al. interviewed six professional haptic designers and found three design themes: the multisensory nature of haptic experiences, a map of the collaborative ecosystem, and the cultural context of haptics \cite{schneider2017haptic}.
The \textbf{inclusivity} and \textbf{customizability} improvements we suggested were similar to their \textbf{EX5: Tailoring and customization}.
They argued that haptic designs need to be tailored to each client’s problem, which was also mentioned in our discussions that we should consider the specific needs of white cane and guide dog users.
In addition, we discussed that users should be allowed to customize their own 3D models, which corresponded to Schneider et al.'s arguments that ``it is “hard for customers to really express what they need and thus, haptic designs need to be tailored'' \cite{schneider2017haptic}.

Moreover, Kim et al. developed a theoretical model of
haptic experience factors including its design parameters, usability requirements, experiential dimensions, and personalization support.
They mentioned the importance of \textbf{timeliness} that haptic experience should have good temporal alignment with other sensory outputs, which corresponded to our discussion that the high-contrast colors should be integrated with haptic feedback to better support low vision users \cite{kim2020defining}.
Additionally, \textbf{customizability} was also highlighted in their work that it is beneficial to allow the users to adjust their haptic settings to match their preferences.
They mentioned that ``if they are in control and they know they’re in control early in the process, they have a better experience.''

\subsection{Potential Assistive Tools Based on Existing Haptic Devices}

Our literature survey focused on a range of haptic assistive tools designed to support people with BLV in their daily lives and professional activities, and the previous subsection listed improvements to these existing tools. 
Beyond these tools, other haptic technologies---commonly employed in fields such as medicine and VR/AR---hold the potential for adaptation into assistive devices. 
These include \textit{pneumatic haptic devices} and \textit{electrical muscle stimulation (EMS) devices} \cite{bouzbib2022can}. 
% These include \textit{pneumatic haptic devices}, \textit{electrotactile haptic devices}, and \textit{electrical muscle stimulation (EMS) devices} \cite{bouzbib2022can}. 
The broader utilization of these haptic tools could enhance support for individuals with BLV, particularly in tasks that are currently undersupported. 

\subsubsection{Pneumatic haptic devices}
\textit{Pneumatic haptic devices} leverage compressed air and airbags to generate pressure and vibration. 
Despite necessitating complex haptic rendering systems, such as air compressors and associated valves, these devices can produce robust forces, reaching up to 9 N.
This capability makes them suitable for applications in medical simulation \cite{stanley2013haptic,herzig2016stiffness,he2015pneuHaptic} and virtual/augmented reality environments \cite{delazio2018force,wang2014footwear}. 
Specifically, the pressure and vibration output from pneumatic haptic devices can enhance experiences for individuals with BLV in gaming and film-watching.
For instance, they can simulate sensations like impacts from opponents or the feeling of rain, as highlighted by Delazio et al. in their work \cite{delazio2018force}.
As discussed in Section \ref{Rarely-used Stimulation Positions For Various Tasks}, haptic feedback holds promise for improving individuals with BLV' film-watching experiences while minimizing distractions from audio descriptions.
However, further research is needed to streamline the overall system, making it more practical for household use.

% \subsubsection{Electrotactile haptic devices}
% \textit{Electrotactile haptic devices} function by generating a local electric current through the skin beneath the electrodes, eliciting electrotactile feedback. 
% Compared to traditional haptic assistive tools like robotic-arm haptic devices and refreshable braille displays/pin arrays, electrotactile devices boast smaller, lighter, and more flexible designs \cite{kaczmarek1991electrotactile,withana2018tacttoo}. 
% This technology offers intensity and direction cues and operates at faster speeds than other haptic feedback modalities, such as thermal feedback \cite{withana2018tacttoo}.
% These attributes make electrotactile feedback well-suited for tasks requiring rapid finger movements, such as understanding graphical information.
% For instance, researchers can employ electrotactile devices to stimulate various positions around the finger, guiding users along the outlines of figures and providing differing stimulation intensities when exploring different areas within the figures.

\subsubsection{EMS devices}
\textit{EMS devices} operate by artificially activating muscles through the application of electric current via electrodes placed on the target muscles.
In contrast to \textit{electrotactile haptic devices}, which stimulate sensory nerves in the skin to generate tactile feedback, EMS devices directly stimulate muscles to induce involuntary contractions, thereby creating kinesthetic feedback \cite{hosono2022feedback}. 
For example, Tanaka et al. developed a haptic device that utilizes EMS to stimulate various positions on the users' neck, forcing the contraction of different neck muscles and enabling head rotation around yaw and pitch axes \cite{tanaka2022electrical}. 
EMS devices can be beneficial for individuals with BLV in tasks requiring kinesthetic feedback, such as graphical information understanding and guidance/navigation. 
In \textit{graphical information understanding}, EMS devices can assist users in navigating through figures and charts by inducing muscle contractions in the arms and fingers. This stimulation prompts users to move their hands along the outlines of the graphical elements, aiding in their exploration and comprehension of the displayed information. 
For \textit{guidance/navigation}, EMS devices can be further developed to offer directional cues by activating neck muscles, thus prompting the user's head to rotate for guidance.

In addition to pneumatic haptic devices and EMS devices, other haptic devices, such as controllable pain feedback devices and dielectric haptic devices, also hold potential applications as haptic assistive tools \cite{jiang2021douleur,chen2022haptag}.  
Our review of these haptic tools illustrates their potential for adaptation to assist individuals with BLV with various tasks. However, future research is needed to explore their efficacy, usability, and acceptance among users with BLV.

\subsection{Relationship of Haptic and Audio Feedback for Various Tasks}

Although our literature survey focused on haptic assistive tools, both haptic and audio feedback serve as the primary sensory channels through which people with BLV perceive the world. 
Many assistive devices employ either or both types of feedback to aid users with BLV, and we found that the application of these feedback types varies across different contexts. 
For \textit{graphical information understanding}, haptic feedback predominates, with audio serving a supplementary role. 
Conversely, in \textit{guidance or navigation} tasks, it is uncommon to use audio and haptic feedback concurrently. 
In \textit{educational and training} settings, audio feedback is the principal medium, and the importance of haptic feedback is generally reduced compared to the other contexts. 

\subsubsection{Relationship of haptic and audio feedback in graphical information understanding}
For \textit{graphical information understanding}, audio cues have been utilized to convey information about various features in figures, such as colors and shapes \cite{shi2016tickers,li2019editing}, specific values in charts like points, bars, and sections \cite{fan2022slide,abu2010multimodal}, and names and descriptions of locations and areas in maps, such as transportation hubs and tourist attractions \cite{gay2021f2t,holloway2018accessible}.
However, studies have indicated that relying solely on audio cues may not facilitate a holistic understanding of figures, charts, and maps \cite{jiang2023understanding}. 
Translating abstract numbers and descriptions into a mental map can be challenging, especially as the complexity of the graphical representations increases. 
In contrast, haptic feedback alone enables users with BLV to establish a mental layout of figures, charts, and maps. 
Nevertheless, without audio descriptions, the connection between the layout and actual data values may be weak.
This is because audio has no spatial constraints and thus can provide temporal information.
In such cases, while screen readers can provide numerical values and detailed descriptions, intuitive comprehension of the structure of figures, charts, and maps relies on haptic cues.
Therefore, integrating both haptic and audio feedback is essential to enhance the comprehension of graphical information, which is supported by previous research \cite{engel2019svgplott,melfi2020understanding}. 

\subsubsection{Relationship of haptic and audio feedback in guidance/navigation}
Audio cues have been infrequently used in conjunction with haptic cues for \textit{guidance/navigation}, but their efficacy in supporting these tasks has often been compared against haptic cues. 
Studies indicate that audio guidance is less favored due to its interference with ambient environmental sounds, which users with BLV rely on to detect approaching pedestrians and vehicles \cite{kayukawa2020guiding,flores2015vibrotactile}. 
For instance, Liu et al. incorporated audio feedback as a supplementary channel to haptic cues to provide descriptive information, such as the distance to the next intersections and the path's curvature \cite{liu2021tactile}. 
However, participants reported confusion between audio and haptic guidance, largely because audio guidance lacks the precision of haptic guidance. 
While haptic cues can be conveyed with precision, such as rotating a servo motor to a specific angle (e.g., fewer than 5 degrees), audio guidance often provides vague instructions like "go straight ahead" instead of specifying directions.
Consequently, users tended to disregard the audio guidance and rely solely on haptic cues, perceiving them as more accurate.

\subsubsection{Relationship of haptic and audio feedback in education/training}
For \textit{education/training}, audio cues are often employed to describe key concepts such as shapes, organs, and landscapes in subjects like mathematics, biology, and geography \cite{murphy2015haptics,crossan2008multimodal}. 
Based on our review, unlike in graphical information tasks, where audio feedback typically serves as a supplementary channel to haptic cues, in educational contexts, audio feedback is often the primary mode of instruction for students with BLV. 
Haptic feedback complements this by presenting various lines, shapes, organ structures, and terrains to facilitate students' understanding of the subject matter.
\rv{Although haptic feedback may not be the dominant modality in educational concepts and settings, it remains a valuable learning modality.} 
Previous studies have highlighted that different learners and tasks have distinct preferences for learning modalities \cite{lodge2016modality}. 
Each modality contributes uniquely to concept development, depending on the learner and the context.
Learning is most effective when the material is delivered through a modality that aligns with the learner’s or task's preference \cite{lodge2016modality,dilkina2013conceptual}.
For example, beyond individual preferences, haptic feedback becomes crucial in specific scenarios, such as teaching complex geometric shapes, where verbal descriptions alone are insufficient to convey the necessary spatial and tactile information effectively.

% However, the necessity for haptic feedback diminishes compared to graphical information tasks, as most concepts can be effectively conveyed through audio feedback alone. 
% Haptic feedback becomes essential only in specific scenarios where complex shapes in geometry, for instance, cannot be adequately explained through verbal descriptions alone. 

\subsection{Limitations of this Systematic Literature Review}
While our work provides an in-depth analysis of existing literature on haptic assistive tools for individuals with BLV, several limitations should be noted. 
First, we concentrated primarily on research prototypes from papers, excluding commercially available assistive tools used by the BLV community. Including studies on commercial products in future work could help identify gaps between theoretical designs and real-world use, potentially guiding improvements in product development.
Additionally, our review only covered papers from the past two decades, overlooking earlier foundational work in the design of assistive tools. Future studies could broaden the scope to include older, seminal research, providing a more comprehensive understanding of the evolution of haptic assistive technologies.
Moreover, our review focused on the HCI perspective, examining tasks and device interactions, such as feedback types and stimulation positions. 
However, researchers from fields like mechanical engineering and \rv{materials} science can also develop assistive devices. 
Future reviews could incorporate papers with more engineering and technical details, providing a broader reference for researchers in these areas.

\section{Conclusion}

To address the literature gap in the comprehensive analysis of haptic assistive tools, feedback types, and on-body stimulation positions across various tasks, we conducted a systematic literature review covering twenty years (2004-2024) of research in the HCI community, drawing from eleven relevant conferences and journals. 
Our review highlighted how different haptic tools, feedback types, and stimulation positions are applied to a range of tasks, and we provide detailed design considerations for their use in specific contexts. 
In addition, we identified and analyzed the limitations of current haptic assistive tools, including challenges related to UX and evaluation methods. Based on this analysis, we proposed future research directions, emphasizing the need for further exploration of haptic tools, feedback types, and stimulation positions to better support a wider variety of tasks. 
We also offered suggestions for enhancing haptic assistive technologies to improve their effectiveness and user experience, ultimately promoting greater accessibility and inclusivity for people with BLV across diverse real-world settings.

\begin{acks}
This work is partially supported by the Guangzhou-HKUST(GZ) Joint Funding Project (No. 2024A03J0617), Guangzhou Science and Technology Program City-University Joint Funding Project (No. 2023A03J0001), the Project of DEGP (No.2023KCXTD042), and Guangdong Provincial Key Lab of Integrated Communication, Sensing and Computation for Ubiquitous Internet of Things (No. 2023B1212010007). 
\end{acks}

\bibliographystyle{ACM-Reference-Format}
\renewcommand{\bibpreamble}{\textit{References marked with • are in the set of reviewed papers.}}
\bibliography{main.bib}

\end{document}